\documentclass[aps,prb,showpacs,twocolumn]{revtex4}
\usepackage{bm,color,amsmath,amssymb,mathrsfs,latexsym,graphicx,psfrag}
\newcommand{\bra}[1]{\left\langle#1\right|}
\newcommand{\ket}[1]{\left|#1\right\rangle}

\newcommand{\beq}{\begin{equation}}
\newcommand{\eneq}{\end{equation}}

\begin{document}

\title{Thin-Torus Limit of Fractional Topological Insulators}

\author{B. Andrei Bernevig${}^1$}
\author{ Nicolas Regnault${}^{1,2}$}
\affiliation{ ${}^1$ Department of Physics, Princeton University, Princeton, NJ 08544} 
\affiliation{$^2$ Laboratoire Pierre Aigrain, ENS and CNRS, 24 rue Lhomond, F-75005 Paris, France}

\begin{abstract} 
We analytically and numerically analyze the one-dimensional "thin-torus" limit of Fractional Topological Insulators in a series of simple models exhibiting exactly flat bands with local hopping. These models are the one-dimensional limit of two dimensional Chern Insulators, and the Hubbard-type interactions projected into their lowest band take particularly simple forms.  By exactly solving the many-body interacting spectrum of these models, we show that, just like in the  Fractional Quantum Hall effect, the zero modes of the thin-torus limit are CDW states of occupation numbers satisfying generalized Pauli principles. As opposed to the FQH where the thin-torus CDW appear in orbital space, in the thin-torus FCI states, the CDW states are in real-space.  We show the counting of the quasihole excitations in the energy spectrum cannot distinguish between a CDW state and a FQH state. However, by exactly computing the entanglement spectrum for the  thin-torus states, we show that it can qualitatively and quantitatively distinguish between a CDW and a fractional topological state such as the FCI. We then discover a previously unknown separation of energy scales of the full FQH energy spectrum in the thin torus limit  and  find that Chern insulator models exhibiting strong isotropic FCI states have  a similar structure in their thin-torus limit spectrum. We close by numerically computing the evolution of energy and entanglement spectra from the thin-torus to the isotropic limit. Our results can also be interpreted as an analysis of one-body, $1$-dimensional topological insulators stabilized by inversion symmetry in the presence of interactions. 
\end{abstract}

\date{\today}

\pacs{73.43.-f, 71.10.Fd, 03.65.Vf, 03.65.Ud}
%\keywords{}

\maketitle

\section{Introduction}

Nontrivial topological insulators are systems  not adiabatically continuable to an atomic limit through a path  obeying some set of symmetries (such as time-reversal, charge conjugation, or a point group symmetry) which characterize the topological class \cite{kitaev-2009AIPC.1134.22K}. When an edge or surface is present, the topological insulator may or may not exhibit gapless edge or surface modes depending on whether the symmetry characterizing its bulk is unbroken or broken by the presence of the surface.  The theoretically proposed \cite{kane-PhysRevLett.95.226801,Bernevig15122006,fu-PhysRevB.76.045302}  and experimentally discovered \cite{Koenig02112007,hsieh-nature2008452} topological insulators are time-reversal invariant, and can be explained mainly through one-body theory, as their parent materials  exhibit weak interactions. However, recent theoretical research has focused on proving the existence of new interacting states of matter, the Fractional Topological Insulators (FTI), which come about when one or more of the topologically nontrivial bands of a one-body nontrivial topological insulator have narrow bandwidth, are fractionally filled, and are subject to interactions \cite{neupert-PhysRevLett.106.236804,sheng-natcommun.2.389,wang-PhysRevLett.107.146803,regnault-PhysRevX.1.021014,neupert-PhysRevB.84.165138,Bernevig-2012PhysRevB.85.075128,wang-PhysRevLett.108.126805,Wu-2012PhysRevB.85.075116,ghaemi-2011arXiv1111.3640G,wang-2012arXiv1204.1697W}. When the interactions overcome the weak band dispersion, new topological phases occur. The parade example of  an FTI is the Fractional Chern Insulator (FCI) : it consists of interacting electrons in a fractionally filled, specially designed flat band (zero bandwidth\cite{sun-PhysRevLett.106.236803,tang-PhysRevLett.106.236802,neupert-PhysRevLett.106.236804,hu-PhysRevB.84.155116}) with nonzero Chern number\cite{haldane-1988PhRvL..61.2015H}. For some specific values of the filling factor, they form a collective state similar to that in the Fractional Quantum Hall effect \cite{neupert-PhysRevLett.106.236804,sheng-natcommun.2.389,regnault-PhysRevX.1.021014} but in the absence of an overall magnetic field (flux) on the sample.

Although several attempts have been made to understand the occurrence of FTI states from first principles, many unresolved issues remain, even in the simplest example of an FTI, that of the FCI. First, it is unclear how to differentiate the FCI state from broken symmetry states such as charge density waves (CDW) with small order parameter: on the lattice, a commensurate CDW state is gapped, could appear at identical momenta as the FQH state, and could be favored by Umklapp processes. The arguments based on the counting of the quasihole excitations in the energy spectrum in the FCI  put forward by us \cite{regnault-PhysRevX.1.021014} could also be applied to the CDW state.  It is  also unclear which interactions favor a FQH state in specific one-body models: while a correlation between smooth Berry curvature and the existence of the FCI state \cite{Parameswaran-2011arXiv1106.4025P,goerbig-2012epjb}  has been observed \cite{Wu-2012PhysRevB.85.075116},  we still do not know how (and in which models) the interplay of local interactions in real space  with the local eigenstate structure in momentum space  leads to robust FCI states.

In the FQH effect on the toroidal geometry, useful information can be extracted from a limiting procedure called the "thin torus limit"\cite{bergholtz-PhysRevLett.94.026802,Bergholtz-2006-04-L04001} in which the aspect ratio of the  torus  is skewed such that one length $|\vec{L}_1|$ of the torus is taken to zero while the product $\vec{L}_1\cdot \vec{L}_2 = 2 \pi N_\phi$ is kept constant, with $N_\phi$ equal the number of fluxes through the system. In this limit, convincing analytical\cite{bergholtz-PhysRevB.77.155308} and numerical arguments\cite{rezayi-PhysRevB.50.17199,seidel-PhysRevLett.95.266405} have shown that the FQH ground-states and quasihole excitations are just the single slater determinant of  orbital occupation numbers exhibiting  the possible  patterns of commensurate CDW  at the filling $N_e/N_\phi= p/q$ with $p,q$ relatively prime and short-range Hamiltonians given by electro-static interaction  (in the case of the Laughlin states) or a generalized $k$-body electrostatic interaction in the case of the $Z_k$ Read-Rezayi (RR) states \cite{read-PhysRevB.59.8084,bergholtz-06prb081308,seidel-PhysRevLett.97.056804,ardonne-2008-04-P04016}. These CDW patterns are on the torus are identical to the "root partitions" patterns \cite{bernevig-PhysRevLett.100.246802}  which index the many-body isotropic FQH states.  As the torus is deformed smoothly from the $|\vec{L}_1| \rightarrow 0$ to the $|\vec{L}_1| = |\vec{L}_2 |$ isotropic limit, it was showed that the CDW states evolve continuously into the isotropic FQH states which exhibit no CDW order parameter.  However, the excitations of the isotropic FQH states can be \emph{counted} in the same way as the excitations of the CDW thin-torus states, even though they exhibit no local order parameter. .

In this paper we perform a similar "thin-torus" analysis of the Fractional Chern Insulators by analyzing the FCI problem on a lattice with $N_x = 1$ sites in the $x$ direction. While on the lattice it is not possible to continuously interpolate between the thin torus and the isotropic limit, subsequent increases of the $N_x$ from $1$ to $2, 3,...$ while keeping $N_x N_y$ constant (when possible) show how the thin torus CDW states evolve into the FCI states in the isotropic limit. We propose a series of simple lattice $1$-D models  which have an exact flat band \emph{with local short-range hoppings}. These models are the $1$-D limit of $2$-D  one-body Chern topological insulators, to which we then add Hubbard-type interactions.  The full spectrum of these interacting  one-dimensional  models in the topologically trivial and nontrivial sides can be obtained exactly for any filling in both the topological side and the atomic limit. The quasihole spectrum, which in this case is made up of zero modes, exhibits the counting of excitations of the FQH state,  even though the state is a CDW, and hence cannot be conclusively used to determine whether an FCI state in the isotropic limit is in the same universality class as the FQH. We however find that the structure of the \emph{full} energy spectrum in the thin torus limit, including excitations, exhibits a separation of scales in the FQH effect into different bands of states distinguishable by their electrostatic energies. We observe the same separation of scales in the thin-torus limit of Chern insulator models with strong isotropic FCI states. The particle entanglement spectrum\cite{sterdyniak-PhysRevLett.106.100405} of the (degenerate) ground-states  can also be obtained exactly in the $1$-D limit. We show that the entanglement spectrum \emph{differs considerably}, both quantitatively and qualitatively,  in the thin torus CDW limit from the isotropic FCI limit, and hence can be used as a diagnostic for the topological nature of the state.  We then numerically analyze several different models which do not have a flat band in the thin torus limit, and show that the structure of their spectrum evolution from the thin-torus to the isotropic limit matches that of the FQH evolution between the same limits. Our results show that a $1$-D one-body topological insulator stabilized by inversion symmetry exhibits CDW states upon the introduction of local interactions, and that the only difference between the trivial and non-trivial Inversion symmetric interacting topological insulators is the range of the interactions when projected to the lower band.

The paper is organized as follows. In section $2$ we describe the general framework for the one-body models, including the flat band limit. We also give a detailed of the two orbital and Kagome models. Section $3$ is devoted to the interacting two orbital model. We show that its one dimensional limit can be handled analytically in the isotropic limit. We prove that at filling factor $\nu=1/3$, the low energy physics is a CDW. In Section $4$, we compute that the particle entanglement spectra of a CDW and we show that it differs dramatically from the one of the Laughlin state. Section $5$ deals with the energy spectrum structure of the FQH in the thin torus limit. We numerically study several FCI models in the thin torus limit and compare them to the FQH case.

\section{The one-body models}

In this paper, a one-body translational invariant $N$-band Bloch Hamiltonian $H = \sum_{{\bf{i}},{\bf{j}}} c_{\bf{i} \alpha}^\dagger h _{\alpha \beta} ({\bf{i}} - {\bf{j}})  c_{{\bf{j}} \beta}$ where the ${\bf{i}}$ and ${\bf{j}}$ denotes the unit cell coordinates, $\alpha, \beta = 1,\ldots, N$ are orbital (and spin, when appropriate) indices, the summation over $\alpha$ and $\beta$ is implicit. The Bloch Hamiltonian is named local if its matrix elements $ h _{\alpha \beta} ({\bf{i}} - {\bf{j}}) $ decay exponentially  with the distance ${\bf{i}} - {\bf{j}}$. Usual tight-binding Hamiltonians are of this form, with  hopping taking place usually between close-neighboring sites.  The Fourrier transform $H= \sum_{{\bf{k}}} c^\dagger_{{\bf{k}} \alpha} h_{\alpha \beta} ({\bf{k}})  c_{{\bf{k}} \beta}$ of such a Hamiltonian contains only limited harmonics of the Bloch momentum $k$. The model can be diagonalized in normal modes $\gamma^{n}({\bf{k}})$  of energy $E_n({\bf{k}})$ where $n = 1, \ldots, N$ is the number of energy levels at each $D$-dimensional (with $D=1,2,3$ experimentally accessible) momentum ${{\bf{k}}} = (2\pi i_1 /N_1, \ldots, 2\pi i_D /N_D)$  where $i_j = 0, \ldots N_j -1$ runs through the number of sites in the $j =1, \ldots, D$'th direction. The diagonalized Hamiltonian reads $H= \sum_{n, {\bf{k}}}  E_n({\bf{k}}) \gamma^{n\dagger}({\bf{k}}) \gamma^{n}({\bf{k}})$ where $\gamma^{n}({\bf{k}}) = u^{n\star}_{\alpha}({\bf{k}}) c_{{\bf{k}} \alpha}$  and $ u^{n}_{\alpha}({\bf{k}}) $ diagonalizes the Bloch Hamiltonian matrix $h_{\alpha \beta} ({\bf{k}})  u^n_{\beta}({\bf{k}}) = E_n({\bf{k}}) u^n_{\alpha}({\bf{k}})$. The projector operator in the $n$'th band reads $P^{(n)}=\gamma^{n\dagger}({\bf{k}}) \ket{0} \bra{0} \gamma^n({\bf{k}})$ with matrix elements $P^{(n)}_{\alpha \beta}({\bf{k}})= u^n_\alpha({\bf{k}}) u^{n \dagger}_\beta({\bf{k}})$. In terms of projectors, the one-body Hamiltonian can be written as $h_{\alpha \beta}({\bf{k}}) = \sum_n E_n({\bf{k}}) P^{(n)}_{\alpha \beta} ({\bf{k}})$.

If the one-body Hamiltonian analyzed is an insulator in which a number $N_1$ of the  "low" energy bands are separated by a full (direct and indirect) gap from the rest "high" energy $N- N_1$ bands, we can then introduce a deformation  $H^{FB}$ of the Hamiltonian $H$ which keeps its eigenstates intact but changes its energies so that they are non-dispersive or Flat Band: 
\beq
H^{FB}  = \sum_{n=1}^{N_1} \sum_{{\bf{k}}  }  - P^{(n)} ({\bf{k}}) + \sum_{n=N_1}^{N- N_1} \sum_{{\bf{k}}}    P^{(n)} ({\bf{k}})
\eneq  The simplified flat-band Hamiltonian and not the full Hamiltonian  is  sufficient to analyze the topological properties of an insulator. For example the Chern number of the occupied bands can be expressed as $C= \sum_{{\bf{k}}} \epsilon_{ij} \text{Tr} [P_G({\bf{k}})  \partial_i P_G ({\bf{k}} ) \partial_i P_G ({\bf{k}})  $ where $   P_G ({\bf{k}} )  = \sum_{n=1}^{N_1} P^{(n)} ({\bf{k}})$  is the projector into all of the occupied bands. The simplified flat-band Hamiltonian is now equivalent in spirit to a Landau Level in the sense that its zero kinetic energy does not compete with with any interaction. At fractional filling of any flat band, the one-body system is a zero-bandwidth metal and exhibits thermodynamic limit degeneracy of the metallic states. Upon introducing repulsive interactions which are projected into the fractionally filled band (thereby artificially rendering gaps infinite - a limit similar to that of projection in the Lowest Landau Level) , two things can happen. In the thermodynamic limit the degeneracy is lifted, but, supposedly in different ways depending on whether the one-body system is topologically trivial or nontrivial.   In the atomic limit, a repulsive interaction will still (most likely) allow for a large degeneracy of states characterized by the configurations of particles on sites minimizing the repulsive local energy. On the side where the one-body Hamiltonian is  topological nontrivial, the hope is that the delocalized nature of the eigenstates will frustrate the local interaction such that  the system will exhibit a finite degeneracy in the thermodynamic limit. A further hope is that this degeneracy is then not due to a symmetry broken gapped state on the lattice such as a CDW, in the thermodynamic and isotropic limit.  

Several Chern insulator models have been studied in presence of strong repulsions : the checkerboard lattice models\cite{sun-PhysRevLett.106.236803,sheng-natcommun.2.389,neupert-PhysRevLett.106.236804,wang-PhysRevLett.107.146803,regnault-PhysRevX.1.021014}, the Haldane honeycomb model\cite{haldane-PhysRevLett.55.2095,Wu-2012PhysRevB.85.075116,wang-PhysRevLett.107.146803}, the ruby lattice model\cite{hu-PhysRevB.84.155116,Wu-2012PhysRevB.85.075116}, the Kagome lattice model\cite{tang-PhysRevLett.106.236802,Wu-2012PhysRevB.85.075116} and the two orbital model on a square lattice\cite{Wu-2012PhysRevB.85.075116}. In this article, we will mostly focus on these two latest examples that we will now describe.

\begin{figure}[h]
\includegraphics[width=0.98\linewidth]{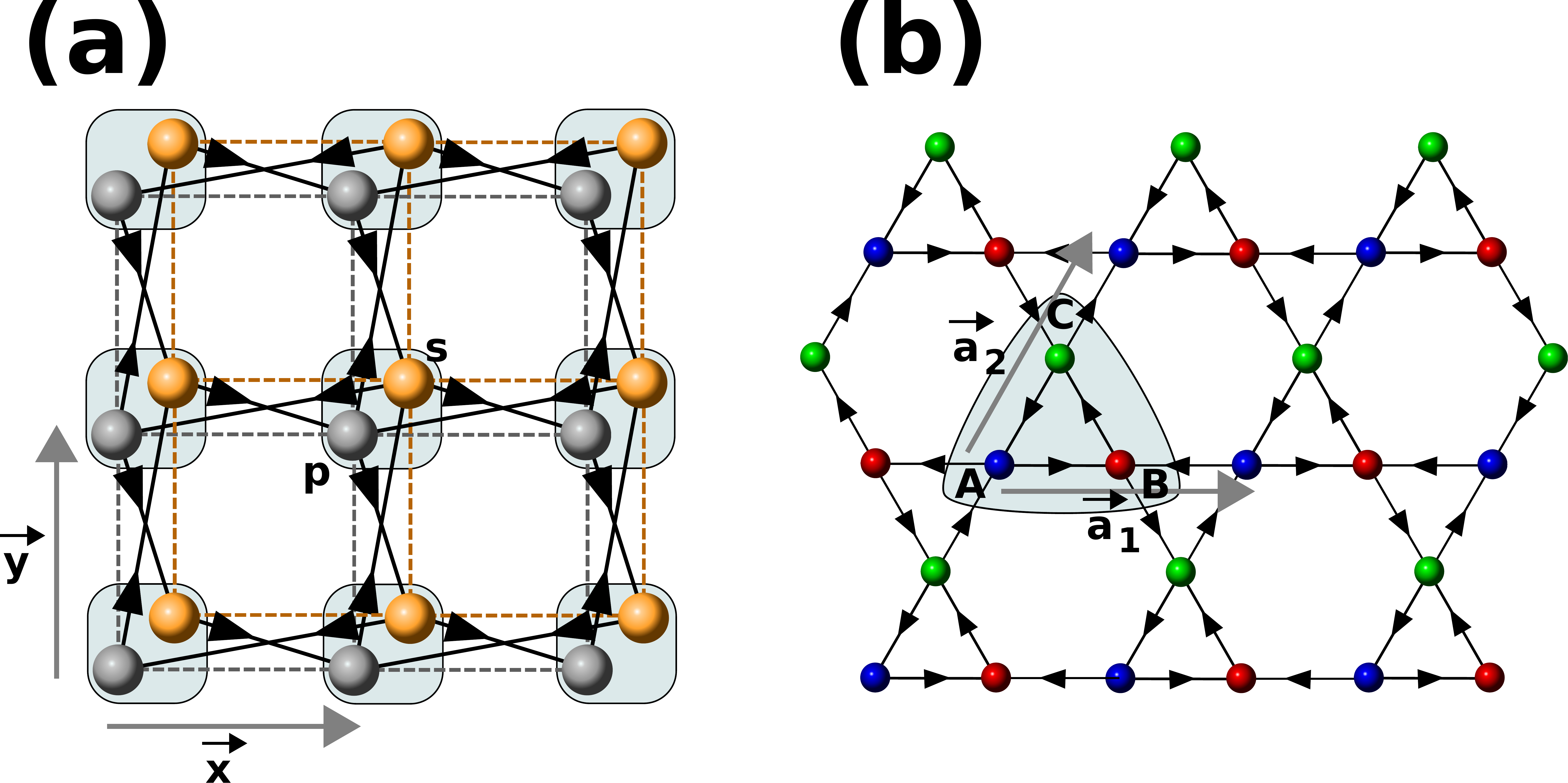}
\caption{\emph{Left panel}:  The two orbital model on the square lattice. The two orbital $s$ and $p$ in each unit cell are depicted by the gold and silver sphere. \emph{Left panel}: the Kagome lattice model made of its the three sublattices $A$, $B$, $C$. The arrows give the sign convention for the phase of the NN hopping term; the hopping amplitude is $\exp (i\varphi)$ in the direction of the arrows. The lattice translation vectors are $\boldsymbol  a_1$ and $\boldsymbol  a_2$.} \label{fig:lattice}
\end{figure}

The simplest example of a Chern insulator model is a two-orbital model on a square lattice with $s$ and $p$ orbitals on site as depicted in Fig. \ref{fig:lattice}a. The nearest neighbor Hamiltonian with a topological transition reads $H= \sum_{{\bf{k}}} (c^{\dagger}_{{\bf{k}},s}, c^\dagger_{{\bf{k}}, p}) h_{\text{s-p}} ({\bf{k}}) (c_{{\bf{k}},s}, c_{{\bf{k}}, p})^T$ :
\beq
h_{\text{s-p}} ({\bf{k}}) = \sum_{i=x,y} -\sin( k_i) \sigma_i - (M- \sum_{i=x,y} \cos (k_i) ) \sigma_z \label{simplesthamiltonianC1}
\eneq where $\sigma_{x,y,z}$ are the three Pauli matrices. This Hamiltonian has $4$ phases, two adiabatically continuable to the atomic limits $M\rightarrow \pm \infty$ with Chern number $C=0$, one phase $0<M<2$ with Chern number $C=-1$ and one phase  $-2<M<0$ with Chern number $C=1$. $M=2,0,-2$ are one-body topological phase transition points where the gap closes in a Dirac fermion mass-changing transition at ${\bf{k}}= (0,0), ((0,\pi)$ and $(\pi,0)), (\pi,\pi)$. In fact, the above Hamiltonian  Eq[\ref{simplesthamiltonianC1}] is the first of a part of a series of Hamiltonians $h^n_{\text{s-p}}$ with $n$'th neighbor hopping: 
\beq
h^n_{\text{s-p}} ({\bf{k}}) = \sum_{i=x,y} \sin (n k_i) \sigma_i + (M- \sum_{i=x,y} \cos(n k_i) ) \sigma_z \label{simplesthamiltonianC2}
\eneq which also exhibit $4$ phases, two adiabatically continuable to the atomic limits $M\rightarrow \pm \infty$ with Chern number $C=0$ phase $0<M<2$ with Chern number $C=-n^2$  and one phase  $-2<M<0$ with Chern number $C=n^2$. $M=2,0,-2$ are one-body topological phase transition points where the gap closes in a Dirac fermion mass-changing transition at ${\bf{k}}= (2 \pi i_1/n, 2 \pi i_2/ n_2), (( 2 \pi i_1/n , 2\pi (i_2 + 1/2)/ n)$ and $(2 \pi( i_1+1/2)/n , 2\pi i_2/n)), (2 \pi( i_1+1/2)/n, 2 \pi( i_2+1/2)/n)$ where $i_1, i_2 = 0, \dots,  n-1$.

The Hamiltonian (\ref{simplesthamiltonianC2}) have a discrete point group symmetry under inversion ${\cal P}= \sigma_z$:
\beq
{\cal P} h^n_{\text{s-p}} ({\bf{k}})  {\cal P}^{-1} =  h^n_{\text{s-p}} (- {\bf{k}}) 
\eneq The inversion eigenvalues $\zeta({\bf{K}})$ of the lower band at the inversion symmetric points ${\bf{K}}= (0,0), (0, \pi), (\pi, 0), (\pi, \pi)$ change in the phase transition change in a phase transition, and the Chern number parity is consistent with the relation $(-1)^C = \prod_{{\bf{K}}} \zeta({\bf{K}})$ \cite{hughes-PhysRevB.83.245132}. 

The other model that we will focus on, especially for the numerical calculations, is the Kagome model\cite{tang-PhysRevLett.106.236802}. It consists of three sublattices $A$, $B$, $C$ spanned by the translation vectors $\boldsymbol a_1$ and $\boldsymbol a_2$ (see Fig \ref{fig:lattice}b). Different tight binding models leading to a nonzero Chern number have been consider in Ref.\cite{tang-PhysRevLett.106.236802}. We consider the simplest case with only the nearest neighbors hopping terms of amplitude $exp(\pm i\varphi)$. The Bloch Hamiltonian reads

\begin{equation}
\begin{split}
h(\boldsymbol k)& = - \left[ \begin{array}{c c c}
				0            &  e^{i\varphi}(1 + e^{-ik_1})      &  e^{-i\varphi}(1 + e^{-ik_2})       \\
				 & 0                   &  e^{i\varphi}(1 + e^{i(k_1 -k_2)}) \\
				{\rm h.c.} &  & 0
				\end{array}
			  \right]
\end{split}
\end{equation}

where $k_1={\boldsymbol k}\cdot{\boldsymbol a_1}$ and $k_2={\boldsymbol k}\cdot{\boldsymbol a_2}$. The phase diagram of this model is determined by the $\varphi$ phase. The gap closes at $\varphi=0,\pm\frac{\pi}{3}$ and away from these points the lowest and highest energy bands have a Chern number equal to $\pm 1$. When looking at the interacting case at filling factor $\nu=1/3$, this system has been shown to host the most robust fermionic Laughlin-like state\cite{Wu-2012PhysRevB.85.075116}. So it is the most natural candidate to consider if one wants to address the question of the thin torus limit. As we will see below, its spectrum in the thin-torus limit is very similar to that of the FQH state without the need for fine-tuning.

\section{The interacting two orbital model}

We now add a density-density interaction   ranging from on-site to the $m$'th nearest neighbor:
\beq
H_{\textbf{U}} = \sum_{{\bf{i}}, {\bf{j}};| {\bf{i}} - {\bf{j}}| \le m} V_{| {\bf{i}} - {\bf{j}}|}  : (n_{{\bf{i}},s} + n_{{\bf{i}},p})(n_{{\bf{j}},s} + n_{{\bf{j}},p}): \label{InteractingHamiltonian} 
\eneq where $n_{{\bf{i}},s}, n_{{\bf{i}},p}$ are the densities $c_{{\bf{i}},s}^\dagger c_{{\bf{i}},s}, c_{{\bf{i}},p}^\dagger c_{{\bf{i}},p}$ of the on-site $s, p$ orbitals and where the term $: \ldots :$ denotes the normal ordering usually performed in FQH calculations \cite{hughes-preparation} where all the $\dagger$ operators get placed to the left with only signs due to permutations. We restrict to $V_{| {\bf{i}} - {\bf{j}}|}$ all positive. Note that for the orbital model this  interaction contains an on-site $n_{{\bf{j}},s} n_{{\bf{j}},p}$ interaction between the two orbitals. The  interacting Hamiltonian is called local if $m$ is finite with $m/N_x \rightarrow 0$ as $N_x\rightarrow \infty$. To work in the lowest band, we now project this interaction to the negative energy band (this corresponds to setting the interaction energy much lower that the artificially modified one-body gap, which is now infinity). In momentum space the projection is local at every $k$ (as the  one-body Hamiltonian is diagonalized in $k$). In real space, however, the interaction now  becomes long-ranged (the interaction will have matrix elements between any two sites on the lattice no matter what the number of sites $N_x, N_y$ is). If the system is gapped, the interaction matrix elements will still decay exponentially for  $| {\bf{i}} - {\bf{j}}| >m$, where $m$ is the original range of the Hamiltonian. Projection of the Hamiltonian to the lowest band has rendered the problem difficult as the eigenstates are strongly dependent on ${\bf{k}}$, especially in the topological nontrivial side, where the "winding" of the eigenstates gives rise to the Chern number.

\subsection{Isotropic Limit: Overview of General Properties}

One limit of the $2D$ problem is easy to analyze: in the atomic limit $M \rightarrow \infty$  (the limit $M \rightarrow - \infty$ is of course identical) the projection to the lowest band fixes the problem to contain only $s$ orbitals (the $p$-orbitals have an infinite energy). The projected interacting Hamiltonian becomes  $P^{(1)} H_{\textbf{U}} P^{(1)} = \sum_{{\bf{i}}, {\bf{j}};| {\bf{i}} - {\bf{j}}| \le m} V_{| {\bf{i}} - {\bf{j}}|}  :n_{{\bf{i}},s}  n_{{\bf{j}},s}: = \sum_{{\bf{i}} \ne  {\bf{j}};| {\bf{i}} - {\bf{j}}| \le m} V_{| {\bf{i}} - {\bf{j}}|}  :n_{{\bf{i}},s}  n_{{\bf{j}},s}:  $. Its spectrum can be determined by combinatoric counting. For example for on-site interaction $m=0$ in the unprojected Hamiltonian, the projected Hamiltonian becomes zero and  the spectrum consists of $
\left( \begin{array}{c}
N_x N_y \\
N_e
\end{array} \right) $ zero modes which count the configurations of $N_e$ fermionic particles in $N_x N_y$ boxes.

\begin{figure}[htb]
\includegraphics[width=0.98\linewidth]{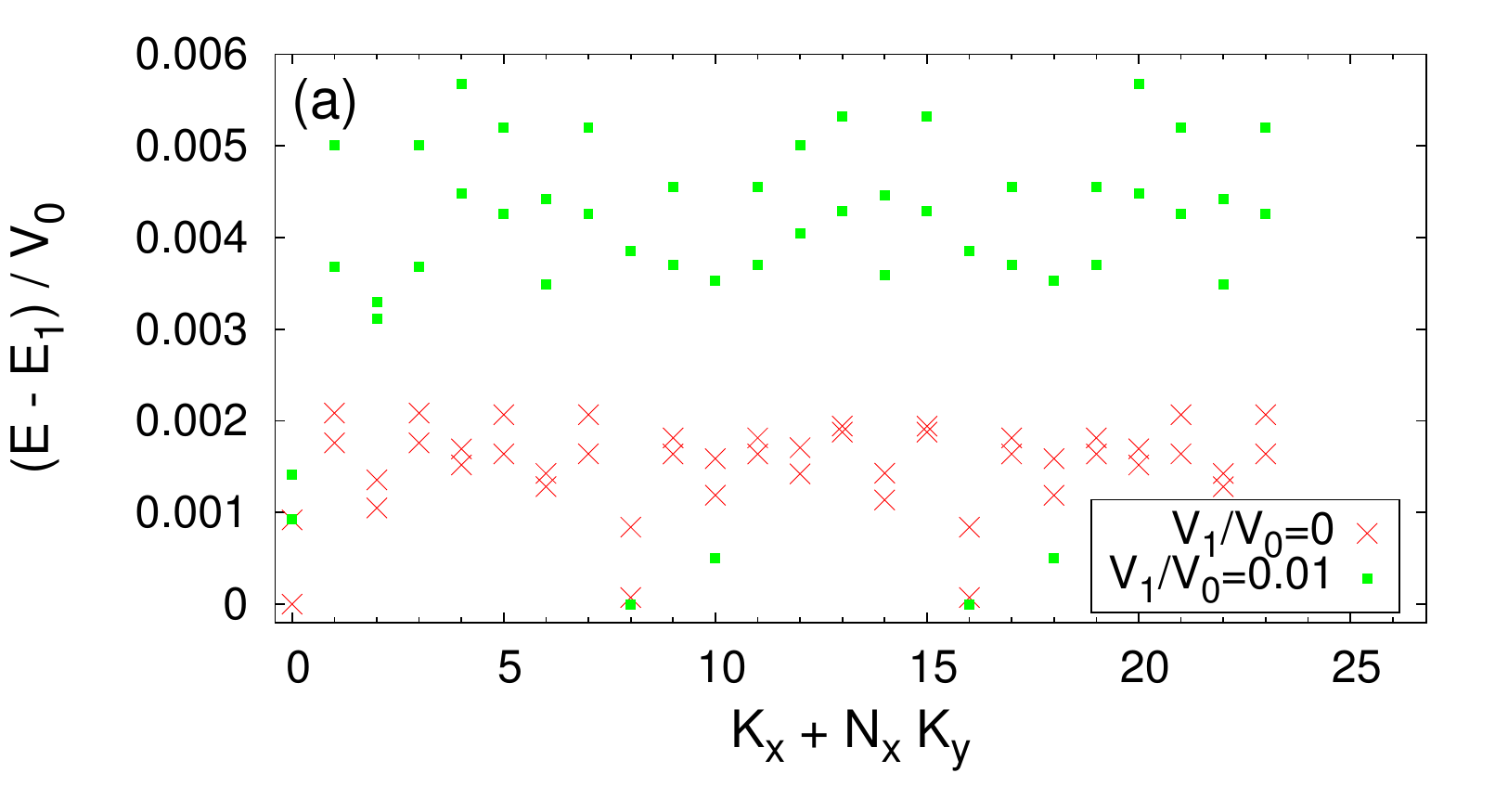}
\includegraphics[width=0.98\linewidth]{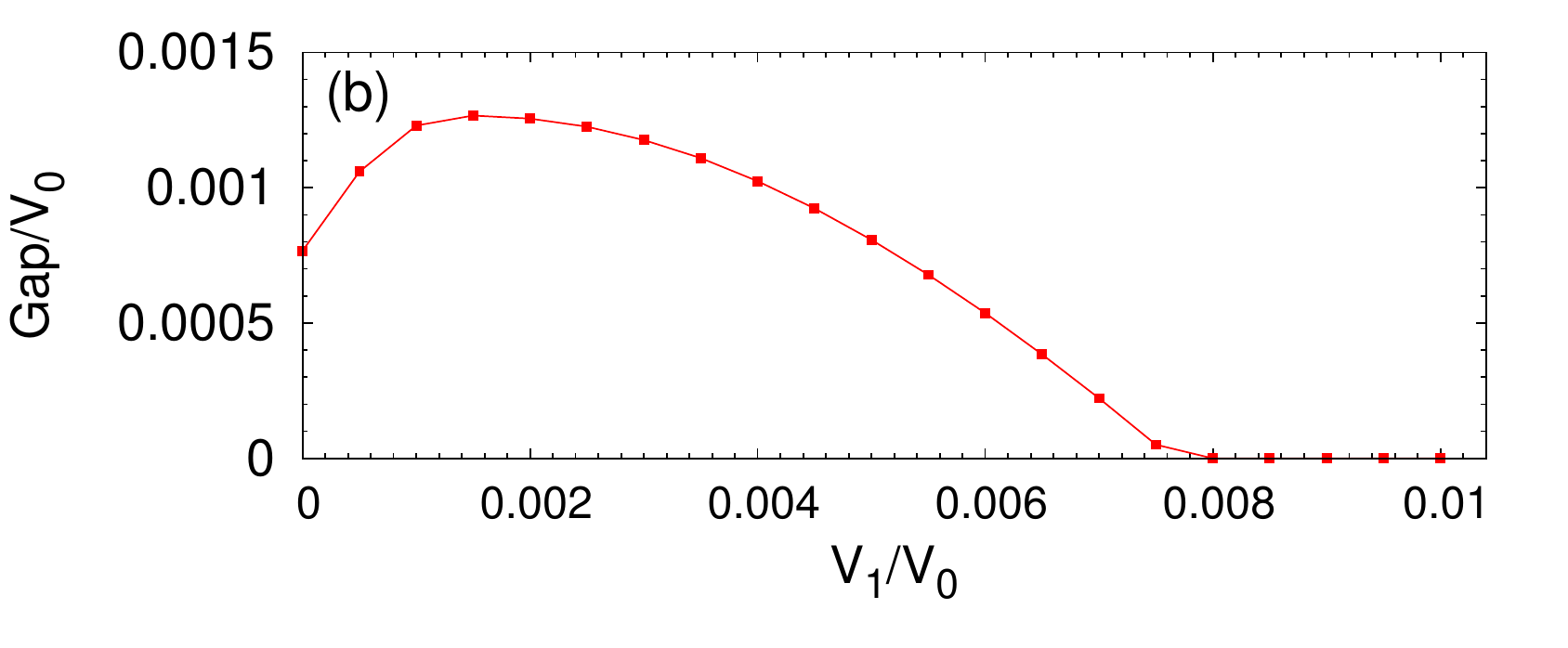}
\caption{{\it Upper panel} : Energy spectrum for $N=8$ fermions on the two orbital model with $N_y=6$, $N_x=4$ and $M=1$. Only the two lowest energies per momentum sector are displayed. Both spectra for $V_1$ and $V_1/V_0=0.01$ are shifted with their respective ground state energy $E_1$. For this system size and this aspect ratio, the Laughlin state $\nu=1/3$ should appear in momentum sectors $(K_x,K_y)=(0,0)$, $(0,2)$ and $(0,4)$. {\it Lower panel} : Gap between the fourth state energy and the third state energy as a function of the nearest neighbor repulsion $V_1$. Until the gap closes, the three lowest energy states have the momenta corresponding to the Laughlin state.}\label{fig:twoorbitalspectrum}
\end{figure}

We now ask what happens in the interacting problem when the  one-body Hamiltonian is driven through a phase transition from the atomic limit $M\rightarrow \infty$ to the $0<M<2$ phase with Chern number nonzero.  With interaction projected into the lowest band,  numerically, at least for $N_x \times N_y = 6\times 4$ and  $6 \times 5$ at $N_e=8, 10$ electrons respectively (filling $1/3$), a $3$-fold degenerate ground--state is observed \cite{Wu-2012PhysRevB.85.075116} with on-site $V_0$ and nearest neighbor $V_1$ interactions, but only for $V_1<<V_0$. For a yet undetermined reason, no gapped degenerate ground-states is obtained for smaller sizes. The observed states are fragile with respect to $V_1$ (see Figs. \ref{fig:twoorbitalspectrum}a and b), though a ratio $V_1/V_0 = 0.0015$ leads to a $40\%$ increase of the gap over the $V_1=0$ value. The fact that the optimal $V_1/V_0$ ratio is small will become important later. 

\subsection{Thin Torus Limit: One Body Model}

 While we cannot solve  the interacting  model analytically in the generic situation, the \emph{thin torus}  limit $N_x =1$ is amenable to theoretical analysis. In this limit the Hamiltonian becomes:
\beq
h^n_{\text{s-p}, 1D} ({\bf{k}}) =- \sin (n k)  \sigma_y - (M-1- \cos(n k) ) \sigma_z \label{simplesthamiltonian1D}
\eneq For $n$ odd and $0<M<2$, this Hamiltonian is a topological insulator stabilized by inversion symmetry with $P=\sigma_z$. For $M>2$, the low energy band has inversion eigenvalue $\zeta(0)= \zeta(\pi) = -1$ for all $n$ (odd and even). At $M=2$, a phase transition  flips the inversion eigenvalues at $k=0$ only for $n$ odd but at both $k=0, \pi$ for $n$ even, and these eigenvalues remain identical in the gapped phase until $M=0$ when  the inversion eigenvalue at $k=\pi$ flips for $n$ odd, while for $n$ even there is no inversion eigenvalue change as the gap closing (for $M=0$) does not take place at an inversion symmetric point in the Brillouin zone. Hence in the case of $n$ odd, the product of eigenvalues $\zeta(0) \zeta(\pi)= -1$ for $0<M<2$ and the system exhibits a charge polarization $\int dk A_k/2\pi = 1/2$  ($\mod 1$) where $A_k =-i  u^{-\star}_{k\alpha} \partial k  u^-_{k \alpha}$ is the Berry phase of the lower band with eigenstate $u^-_k$.  The value of the charge polarization is fixed by inversion symmetry and is an invariant of the one-body topological phase - it cannot be changed unless a phase transition occurs. By contrast, the $n$ even models exhibit unit charge polarization and are trivial under the inversion symmetry charge polarization classification. When the system is cut in half, the cut (or edge) breaks inversion symmetry, and the system does not need to exhibit mid-gap modes in its open-boundary energy spectrum (although the Hamiltonian (\ref{simplesthamiltonian1D}) does exhibit these modes, one on each cut, due to the absence of a kinetic term $\epsilon(k)$ proportional to the identity matrix,  which we neglected for aesthetic purposes but which is allowed by inversion symmetry). The system however exhibits mid-gap states in its entanglement spectrum \cite{prodan-PhysRevLett.105.115501}, another way in which it reveals its topological nature. For the $2$-dimensional models we chose, their thin torus limit is, for $n$ odd a topological insulator with inversion symmetry, and our analysis also reveals the fate of an inversion symmetric topological insulator in $1$ dimension (we note that one needs a symmetry in $1$D to define a topological insulator). Our results, however, will be valid when inversion symmetry is absent, and we focus on the thin-torus limit of the FCI rather than on the $1$-D interacting inversion symmetric topological insulator. 

We  further note a special point of the model in Eq[\ref{simplesthamiltonian1D}]: the point for which the gap is maximal is also the point at which the Hamiltonian has an exact flat-band without the need for projection. Indeed, for $M=1$, the energy spectrum of the Hamiltonian is $E_\pm=\pm 1$. Note that  only the thin torus limit has a fully flat band at $M=1$ - the isotropic limit has dispersion in the second direction. Generically, however, it is true that the thin torus limit Hamiltonian has flatter bands than its isotropic counterpart. The fact that the Hamiltonian $h^n_{\text{s-p}, 1D} ({\bf{k}})$ in  Eq[\ref{simplesthamiltonian1D}] has a flat band with only short range ($n$'th neighbor maximum) hopping is revealing as it means that the projectors into the lowest energy band are short ranged and hence amenable to analytic derivation.  The $M=1$ Hamiltonian is easiest to understand in real space:
\begin{eqnarray}
& H_{M=1}^n=\nonumber \\ &-\sum_{{\bf{k}}} (c^{\dagger}_{{\bf{k}},s}, c^\dagger_{{\bf{k}}, p}) ( \sin (n k)  \sigma_y + \cos(n k)  \sigma_z)  (c_{{\bf{k}},s}, c_{{\bf{k}}, p})^T= \nonumber \\ & = -\sum_{j=1}^{N_x} a_j^\dagger b_{j+n} + h.c. \label{heegerschrieffersu}
\end{eqnarray} where the new fermionic operators $a_j,b_j$ on site $j$ are a combination of the $s,p$ on-site orbitals:
\beq
a_j = \frac{1}{\sqrt{2}} (c_{js} - c_{jp}), \;\;\;\; b_j = \frac{1}{\sqrt{2}} (c_{js} + c_{jp})
\eneq We note that the Hamiltonian (\ref{heegerschrieffersu}) makes sense only if $N_y/n \in {\cal{Z}}$. By contrast, the trivial atomic limit Hamiltonian reads:
\beq
H_{M \rightarrow \infty}^n= - M \sum_{j=1}^{N_x} a_j^\dagger b_j + h.c.
\eneq Note that the expression of the Hamiltonian in terms of the new fermionic operators $a_j, b_j$  is just the rotated Hamiltonian Eq[\ref{simplesthamiltonian1D}]: to the $ \sin (n k)  \sigma_y + \cos(n k)  \sigma_x$ basis.  A simple decomposition into normal modes is also more revealing in real space:
\begin{eqnarray}
 H_{M=1}^n &=   \sum_{j=1}^{N_x}& -\frac{a_j^\dagger + b_{j+n}^\dagger}{\sqrt{2}}\frac{a_j + b_{j+n}}{\sqrt{2}}\\
&&+ \frac{a_j^\dagger - b_{j+n}^\dagger}{\sqrt{2}}\frac{a_j - b_{j+n}}{\sqrt{2}} \nonumber
\end{eqnarray} and 
\begin{eqnarray}
& H_{M\rightarrow \infty}^n =   \sum_{j=1}^{N_x} -\frac{a_j^\dagger + b_{j+n}^\dagger}{\sqrt{2}}\frac{a_j + b_{j}}{\sqrt{2}}+ \frac{a_j^\dagger - b_{j}^\dagger}{\sqrt{2}}\frac{a_j - b_{j}}{\sqrt{2}}
\end{eqnarray} The normal mode of the lowest band is hence $\gamma_j= (a_j + b_{j+n})/\sqrt{2}$  on the nontrivial and $\gamma_j= (a_j + b_j)/\sqrt{2}$ on the trivial side respectively. Hence the nontrivial side is a generalized Su- Schrieffer-Heeger model with perfect $n+1$-merization. The ground-states of the two limits are $\psi^n_{M=1} = \prod_{j=1}^{N_x}  (a_j + b_{j+n})\ket{0}$  containing dimerizations between site $j$ and site $j+n$ on the nontrivial site and  $\psi^n_{M\rightarrow\infty} = \prod_{j=1}^{N_x}  (a_j + b_{j}) \ket{0}$ or the $s$-wave orbital on each site $j$ occupied. The most well-known model of the non-trivial side is the $n=1$ model which contains a dimerization between site $j$ and $j+1$, with the center of charge mid-bond. A cut on the bond cuts the charge center. The $n$-generalization of this model exhibits expectation values (dimerization)  of operators on site $j$ and $j+n$. Cutting one bond cuts through $n$ of the dimerized lines. 

\subsection{Thin Torus Limit: Projected Interaction}

We now try to project the interacting Hamiltonian to the lowest bands.  In general, in the isotropic limit, the projection to the lowest band has several consequences. First, even though the interacting Hamiltonian Eq[\ref{InteractingHamiltonian}] is density-density, projection to the lowest band will make it unwritable in terms of just density operators on site. Second, it will introduce nonlocal long-range couplings that run through the length of the lattice. This makes the Hamiltonian very difficult to analytically investigate.  The current $1$-D flat band case is different. Due to the fact that the projection is local, the projected Hamiltonian is also local.   The projection to the lowest band of any interacting Hamiltonian is now easy to do and accomplished by taking $P^{(1)}_n a_j P^{(1)}_n= \gamma_j/\sqrt{2}$, $P^{(1)}_n b_j P^{(1)}_n = \gamma_{j-n}/\sqrt{2}$ in the nontrivial phase and   $P^{(1)}_n a_j P^{(1)}_n= \gamma_j/\sqrt{2}$, $P^{(1)}_n b_j P^{(1)}_n = \gamma_{j}/\sqrt{2}$ on the trivial side, where $P^{(1)}_n$ is  the projector into the lower band for the Hamiltonian $h_{\text{s-p} 1D}^n({\bf{k}})$ at the points $M=1$, $M \rightarrow \infty$ (nontrivial side and atomic limit).   Hence the projected initial operators  of the $s$ and $p$ orbitals in the nontrivial phase read $P^{(1)}_n c_{js} P^{(1)}_n = (\gamma_j + \gamma_{j-n})/\sqrt{2}$ and $P^{(1)}_n c_{jp} P^{(1)}_n = (\gamma_{j-n} - \gamma_{j})/\sqrt{2}$, while in the  atomic limit they are  $P^{(1)}_nc_{js} P^{(1)}_n  = \gamma_j $ and $P^{(1)}_n c_{jp} P^{(1)}_n  =0 $.  A density-density term interacting between $r$ sites in the  original Hamiltonian  Eq[\ref{InteractingHamiltonian}]  is $n_{j+r\alpha }, n_{j \beta}$. When normal ordered and projected to the lowest band becomes, we obtain for $n\ne r$ :
\begin{eqnarray}
& &:\sum_{j \alpha, \beta} P^{(1)}_n n_{j+r, \alpha} n_{j, \beta} P^{(1)}_n:   \nonumber \\ 
& =&\sum_j   2\tilde{n}_{j+r} \tilde{n}_j + \tilde{n}_{j+r+n}  \tilde{n}_j + \tilde{n}_{j+r-n} \tilde{n}_j
\end{eqnarray} 
for $n=r>0$ :
\begin{eqnarray}
& &:\sum_{j \alpha, \beta} P^r n_{j+r, \alpha} n_{j, \beta} P^r: \nonumber\\
& =& \sum_j   2\tilde{n}_{j+r} \tilde{n}_j + \tilde{n}_{j+2 r }  \tilde{n}_j \end{eqnarray} 
and  for $r=0$:
\begin{eqnarray}
 :\sum_{j \alpha, \beta} P^{(1)}_n n_{j, \alpha} n_{j, \beta} P^{(1)}_n: &=&  \sum_j   \tilde{n}_{j+n} \tilde{n}_j \end{eqnarray}
 where $\tilde{n}_j= \gamma_j^\dagger \gamma_j$ is now the density operator of the normal modes $\gamma_j$. One sees that in the topologically ordered side, an interaction between the particle densities $r$ sites away becomes, when projected to the flat band of $h_{\text{s-p} 1D}^n({\bf{k}})$,  an interaction at $r+ n$ sites away.  The only result of the projection is to increase the range of the interaction.   By contrast, on the trivial side, the projected Hamiltonian is zero for $r=0$ and $\tilde{n}_{j+r} \tilde{n}_j$ for $r\ne 0$ - in other words, the range of the Hamiltonian remains unchanged.  The density-density projected Hamiltonians act on the Hilbert space $\prod_{i=1}^{N_e} \gamma_{j_i}^\dagger \ket{0}$, and solving them becomes as simple problem of solving for configurations minimizing a hard-core density-density interaction with no kinetic energy. The interactions can now be tuned to obtain the correct degeneracy of FCI states in the thin torus limit, which is known from the FQH  thin torus limit. In that case, the Hamiltonians are density -density in momentum space (as we will see in the next sections)  and its degenerate ground-states are zero-mode single Slater determinants of CDW-patterns in momentum space \cite{bergholtz-PhysRevLett.94.026802,Bergholtz-2006-04-L04001,bergholtz-PhysRevB.77.155308}. It is now clear that all the FCI thin torus states are commensurate CDW states but in real space. A similar analysis can obviously be extended to higher-density interactions such as those necessary to give a Moore-Read\cite{Moore1991362} or RR \cite{read-PhysRevB.59.8084} state. For bands with some dispersion, which are not flat-band from the get-go, other longer range terms of the Hamiltonian are generated, but we expect that the major properties of the spectrum are maintained.
 
\subsection{The $\nu=1/3$ case}
 
We now  particularize to the case of electron filling  $1/3$ and Chern number $1$ in the isotropic limit. This means we are looking at the $h_{\text{s-p}}^n$ Hamiltonian which for $M=1$ and $N_x=1$ has the thin torus limit $h_{\text{s-p} 1D}^1$ with a flat band. To obtain a $3$-fold exact degeneracy (same as that of the Laughlin FQH state) in the thin torus limit it is now obvious one needs a combination of on-site (inter-orbital) and nearest neighbor density-density interaction:
\beq
H_{\textbf{1/3}} = V_0 \sum_{{\bf{i}}}   n_{{\bf{i}},s} n_{{\bf{i}},p}  + V_1  \sum_{ \langle {\bf{i}}, {\bf{j}}  \rangle } (n_{{\bf{i}},s}+ n_{{\bf{i}},p}  )(n_{{\bf{j}},s}+ n_{{\bf{j}},p})    \label{InteractingHamiltonianIsotropic} 
\eneq where normal ordering is no longer an issue. The relative scales $V_1/V_0$ for which a FQH state is seen in the isotropic limit will be borrowed from the analysis of the FQH problem in the thin torus limit, to be presented soon. When projected to the lowest band, in the topologically nontrivial flat-band case $M=1$, this Hamiltonian becomes:
\beq
P^{(1)}_1 H_{\textbf{1/3}} P^{(1)}_1 = (V_0+ 2V_1) \sum_{  \langle {\bf{i}}, {\bf{j}}  \rangle }   \tilde{n}_{{\bf{i}} } \tilde{n}_{{\bf{j}} }  + V_1  \sum_{ \langle \langle {\bf{i}}, {\bf{j}} \rangle  \rangle } \tilde{n}_{{\bf{i}} } \tilde{n}_{{\bf{j}} }    \label{InteractingHamiltonianprojectednontrivial} 
\eneq while in the atomic limit $M\rightarrow \infty$ the Hamiltonian becomes:
\beq
P^{(1)}_1 H_{\textbf{1/3}} P^{(1)}_1 = V_1  \sum_{ \langle {\bf{i}}, {\bf{j}}   \rangle } \tilde{n}_{{\bf{i}} } \tilde{n}_{{\bf{j}} }    \label{InteractingHamiltonianprojectedtrivial} 
\eneq Notice that the difference between the projected Hamiltonians in the non-trivial and trivial phases is the range of the interaction. At a certain filling, the non-trivial Hamiltonian will have far less zero modes than the trivial one.

The spectrum of \ref{InteractingHamiltonianprojectednontrivial}  is easily obtained for $V_0,V_1>0$. At filling $N_e/N_y=1/3$ ($N_y$ is the number of sites in the $1D$ lattice), the Hamiltonian has a $3$-fold degenerate groundstate which is a charge density wave 
\beq
\ket{1} =\prod_{i=1}^{N_e} \gamma^\dagger_{3 i} \ket{0},  \ket{2} = \prod_{i=1}^{N_e} \gamma^\dagger_{3 i+1} \ket{0}, \; \ket{3} = \prod_{i=1}^{N_e} \gamma^\dagger_{3 i+2} \ket{0}, \label{thintorusCDWstates}
\eneq These are the thin-torus CDW states in which the Laughlin state evolves.  At lower filling $N_e/N_y <1/3$ there are many zero modes given by the number of ways of putting $N_e$ electrons in $N_y$ sites without any two electrons coming closer than $3$-sites together  (in the FQH literature, as mentioned below, this is the $(1,3)$ Pauli principle, though there it appears in orbital space).

The excitation spectrum of [\ref{InteractingHamiltonianprojectednontrivial}] can also be easily obtained exactly, although for the case when $V_0,V_1$ are of the same order of magnitude, it involves a certain amount of book-keeping. However if, as in the FQH thin torus case analyzed in the next section, $V_0>>V_1$ (what we need is that $V_0>> (N_e-1) V_1$, which is more than satisfied in the thin-torus FQH), the excitation spectrum separates into two bands, a phenomenon which we call separation of scales: one band, between energy $V_1$ and $(N_e-1) V_1$ is the band of wavefunctions which are Slater determinants  $\prod_{i=1}^{N_e} \gamma^\dagger_{ j_i} \ket{0}$ such that the distance between any two sites $3\ge |j_{i_1} - j_{i_2}| >1$ and at least one pair of $j$'s have distance $2$ (i.e. not all $j$'s be separated by a distance $3$, as those would be zero modes). In the FQH, these configurations of Slater determinants satisfy a $(1,2)$ Pauli principle which says that there can be no more than $1$ particle in $2$ consecutive sites. Notice that the $(1,3)$ Pauli principle is a sub-principle  of the $(1,2)$ Pauli principle, and hence the number of excitations with energies between $V_1$ and $(N_e-1)V_1$ is equal to the number of configurations of  $N_e$ electrons in $N_y$ sites satisfying a $(1,2)$ Pauli principle minus the number configurations satisfying a $(1,3)$ Pauli principle. The last band of excitations, from energy $V_0+ 2V_1$ to $(N_e-1)(V_0+ 2V_1)$ is that of all fermionic configurations (Pauli principle $(1,1)$) minus the $(1,2)$ and $(1,3)$ configurations. This is the nontrivial $M=1$ case. The current explanation of the spectrum can be extended to other Hamiltonians such as  those that give the RR states - with similar conclusions - in the thin torus limit the zero modes are given by the CDW patterns \cite{bergholtz-06prb081308,ardonne-2008-04-P04016,seidel-PhysRevLett.97.056804} or the root partition patterns \cite{bernevig-PhysRevLett.100.246802} of the respective RR state in real space. The atomic limit $M\rightarrow \infty$ has a spectrum made out of all $1, 2$ excitations as zero modes (it has a size-dependent and increasing number of zero-modes) and the remaining $(1,1)$ excitations at energies $V_1$ to $(N_e-1)V_1$ (not $V_0$). Note that our analysis of the current $1$-D system can also be interpreted as  the analysis of a topological insulator stabilized by inversion symmetry, although the ground-states we obtain are always charge density waves.

The thin torus limit of this exactly solvable model teaches us several things: first, the FCI states in this case are exactly degenerate zero modes trivial commensurate CDW state in real space ($3$ states for Laughlin $1/3$ state). Second, the number of zero mode excitations at lower filling than $1/3$ (the number of "quasiholes") is the same as in the isotropic FQH, even though in the thin torus limit they are excitations of the trivial CDW state (they are single Slater determinants). This last point shows that it is not possible to infer the character of a FQH state just by counting excitations in the energy spectrum - a CDW state would also have the same excitations. Third, in order to mimic the behavior of the FQH spectrum (obtained below), the interaction has to be such that $V_0>>(N_e-1)V_1$. Indeed, in the two-orbital model \ref{simplesthamiltonianC1} turns out to only have a $3$-fold degenerate state separated by a gap in the isotropic limit for $V_0>>>(N_e-1) V_1$ as we have seen in the numerical data  in Figs. \ref{fig:twoorbitalspectrum}a and  \ref{fig:twoorbitalspectrum}b. In this case, the thin torus limit spectrum of the FCI and FQH is very similar (including the separation of scales mentioned above and in the FQH section below).

\section{Entanglement spectrum}

The counting of the quasihole subspace cannot be used as a clear diagnostic of the FCI state in the isotropic limit, as the CDW state (which is actually the ground state in the thin-torus limit) has excitations  obeying identical counting to that of the FQH quasiholes. How can we then clearly identify that the state in the isotropic limit is a topological (FQH) state and not a CDW (without computing order parameters)? We now show, by direct computation, that the entanglement spectrum\cite{li-08prl010504} can distinguish between the CDW state and the isotropic FQH/FCI topological state.  We use the particle entanglement spectrum\cite{sterdyniak-PhysRevLett.106.100405} (PES). For a $d$-fold degenerate state $\{|\psi_i>\}$, we consider the density matrix $\rho=\frac{1}{d}\sum_{i=1}^{d}|\psi_i><\psi_i|$. We divide the $N$ particles into two groups $A$ and $B$ with respectively $N_A$ and $N_B$ particles. Tracing out on the particles that belong to $B$, we compute the reduced density matrix $\rho_a={\rm Tr}_B \rho$. This operation preserves the geometrical symmetries of the original state, so we can label the eigenvalues $\exp(-\xi)$ of $\rho_A$ by their corresponding momenta, but generically we can also look at all momenta sectors without separating quantum numbers (which is what we will do here). The entanglement spectrum is just the $\xi$'s (generally called energies) plotted as a function of the momentum. Generically, for a contact (such as pseudopotential) interaction, the number of non zero eigenvalues in $\rho_A$ is bounded from above by the number of zero-modes of the $N_A$ particles in the large Hilbert space of the original state. For example, for a pseudopotential interaction in the FQH effect for $N_e$ electrons in $N_\phi$ fluxes, the maximum number of nonzero eigenvalues of $\rho_A$ is bounded by the number of zero modes of $N_A$ particles in $N_\phi$ fluxes subject to the same pseudopotential interaction. For FQH model states such as the Laughlin state, the spectrum saturates the bound: the number of non zero eigenvalues in $\rho_A$ matches exactly the number of quasihole states for $N_A$ particles and the same number of flux quanta as the original state. The quasihole counting is characteristic of each model, thus the PES acts as a fingerprint. As was shown in many previous papers\cite{sterdyniak-PhysRevLett.106.100405,regnault-PhysRevX.1.021014,Bernevig-2012PhysRevB.85.075128,sterdyniak-PhysRevB.85.125308,dubail-PhysRevB.85.115321,rodriguez-2011arXiv1111.3634R}, the isotropic FQH/FCI state has a low-entanglement energy spectrum whose levels saturate the quasihole bound.

We now ask if the entanglement spectrum of the CDW ground-states in the thin torus limit is different than the one  in the isotropic limit. For the case of our flat-band model (and similarly for the FQH thin torus limit presented below), we can compute the entanglement spectrum exactly. To make contact with the FQH, we will first go to momentum space and then compute the spectrum in that basis (although we will not solve per momentum quantum number).  Let us show as an example, the PES for the $\nu=1/3$ state in the case when the one-body Hamiltonian has $n=1$: $h_{M=1}^{n=1}(k)= \cos(k) \sigma_z+ \sin(k) \sigma_x$. There are $3$ CDW states $\ket{1}$, $\ket{2}$ and $\ket{3}$ which are zero modes at filling $1/3$ of the projected interacting Hamiltonian with on-site and nearest neighbor interaction (which in the projected basis of operators $\gamma_j$ becomes a nearest neighbor and next nearest neighbor interaction), written down in Eq[\ref{thintorusCDWstates}]. We can form momentum combinations of these states:
\beq
\ket{{\bf{K}}} = \frac{1}{\sqrt{3}}\sum_{j=1}^3 e^{i K j } \ket{j}
\eneq 
with ${\bf{K}}=\frac{2\pi}{3} K + \frac{\pi}{3} ((N_e+1){\rm mod} 2)$ and $K$ being $0$,$1$ or $2$. The FQH entanglement spectrum is formed by taking the superposition of the density matrix of the states at the $3$ $K$ momenta (so that translational invariance is kept). A summation over the values of $K$ leads to the identity:
\beq
\frac{1}{3}\sum_{{\bf{K}}}\ket{{\bf{K}}} \bra{{\bf{K}}} = \frac{1}{3} \sum_{j=1}^3 \ket{j} \bra{j}
\eneq and the density matrix has the same expression in both momentum and real space. Tracing out $N_e- N_A$ particles out of the $N_e$ electron states, we remain with a density matrix for $N_A$ particles on the sites of the lattice. Since the $3$ states $\ket{j}$ are orthogonal CDW states with electrons fixed at lattice positions, the entanglement spectrum breaks up into $3$ sectors, and has degenerate eigenvalues (this statement is also true for the thin torus limit  abelian state whose degeneracy is just a center of mass degeneracy- for example for the $1/m$ Laughlin state, the spectrum breaks up into $m$ sectors and is completely degenerate). The number of states per each sector is just the number of ways of choosing $N_A$ electrons out of the initial $N_e$ electrons,  $\left( \begin{array}{c}
N_e\\
N_A
\end{array} \right) $. The total number of eigenvalues is:
\beq
{\cal N}^{(N_A,N_e)}_{CDW}= 3 \left( \begin{array}{c}
N_e\\
N_A
\end{array} \right) \label{CDWcountringontorus}
\eneq All the entanglement energies are exactly degenerate at $1/{\cal N}^{(N_A,N_e)}_{CDW}$. This number is significantly smaller than the number of  eigenvalues that one would see for the Laughlin state, which is 
\beq
{\cal N}^{(N_A,N_e)}_{Lgh}=\frac{3N_e (3N_e  - 2 N_A-1)!}{N_A! (3 N_e - 3N_A)!} \label{quasiholecountringontorus}
\eneq
In the thermodynamic limit, the difference between the two counting diverges. The CDW counting is just the counting of $(1,3)$ configurations of $N_A$ particles that \emph{are already present in the  CDW groundstate} of the system while the FQH counting is that of all $(1,3)$ configurations of $N_A$ particles in $3 N_e$ orbitals. Hence the entanglement spectrum of the CDW does \emph{not} saturate the bound by a large value, and has far fewer levels than the entanglement spectrum of the FQH state.

\begin{figure}[htb]
\includegraphics[width=0.98\linewidth]{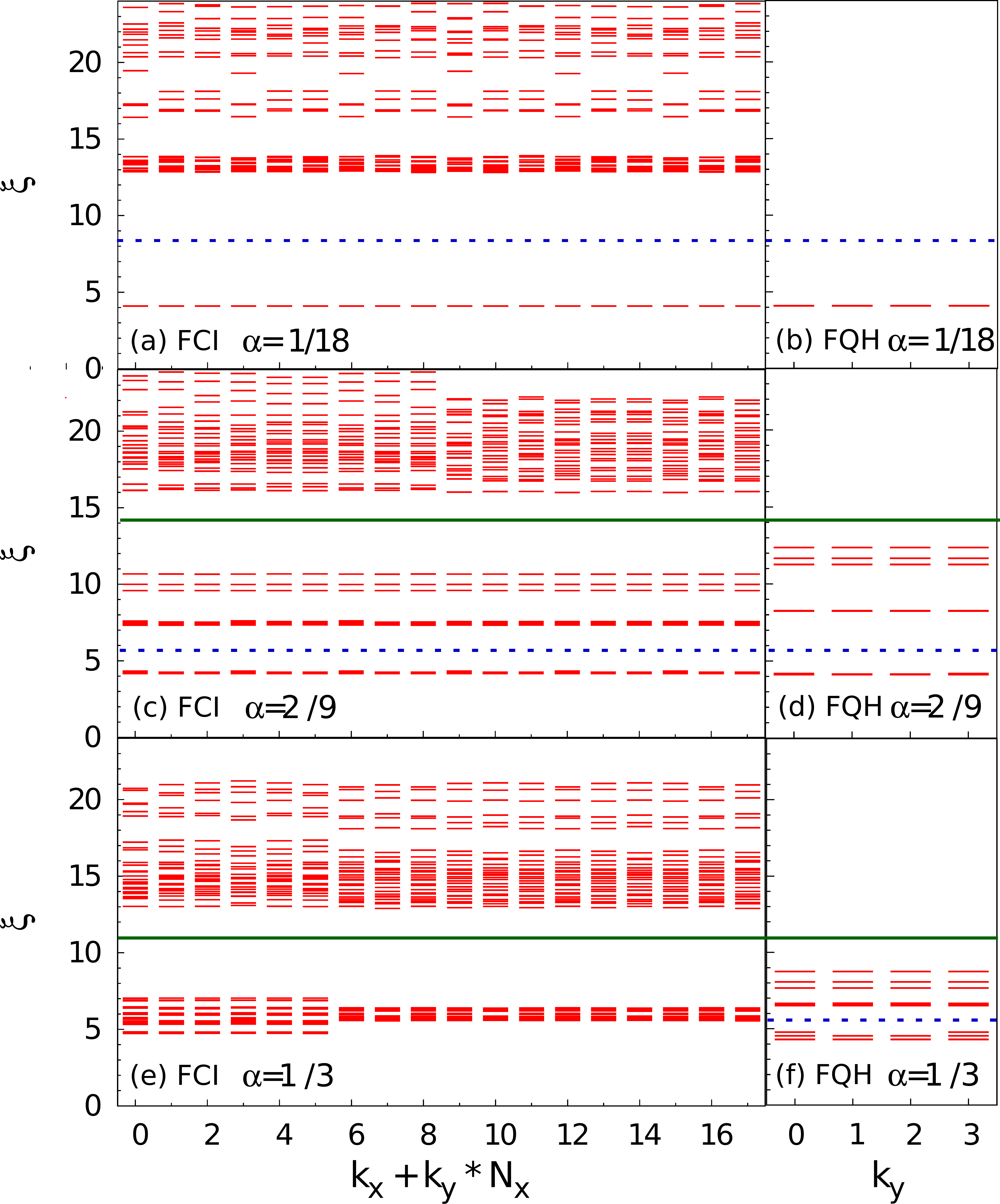}
\caption{PES built from the three low energy states for $N=6$ fermions, keeping $N_A=3$ particles. PES are shown for the FCI with $N_x=18, N_y=1$ (fig. a), $N_x=9, N_y=2$ (fig. c), $N_x=6, N_y=3$ (fig. e) and the FQH on a torus with aspect ratios $\alpha=1/18$ (fig. b), $\alpha=2/9$ (fig. d), $\alpha=1/3$ (fig. f). The state counting below the solid green line matches the $(1,3)$ Laughlin quasihole counting in  Eq[\ref{quasiholecountringontorus}], the state counting below the dashed blue line matches the CDW counting  in Eq[\ref{CDWcountringontorus}]. Notice that the PES of the two orbital model in the isotropic case and of the FQH in the thin torus limit are identical.}\label{entspectrum}
\end{figure}

For models without a perfect flat-band limit, the main features of the entanglement spectrum remain. The CDW-like ground-states will have different entanglement spectrum than the topological state. Going from the thin torus to the isotropic limit, the remaining levels in the entanglement spectrum should come down and mix with the CDW-like levels to give the full number of FQH entanglement levels in Eq[\ref{quasiholecountringontorus}]. To show this, we plot the evolution of the entanglement spectrum of the Kagome model from the thin torus limit to the isotropic limit in Fig. \ref{entspectrum}. In the thin torus limit (Fig. \ref{entspectrum}a), the number of entanglement energies is exactly given by ${\cal N}^{(N_A,N_e)}_{CDW}$ and the energies are almost all degenerate. The repulsive interaction that we have used is the nearest neighbor interaction\cite{Wu-2012PhysRevB.85.075116}. Fig. \ref{entspectrum}c and \ref{entspectrum}e show how the remaining levels mix with the thin torus CDW levels as the aspect ratio is made isotropic. Notice that there is a clear entanglement gap between the low energy physics driven by the Laughlin state and higher energy states. Hence the entanglement spectrum can be used to differentiate between the CDW and the FQH state. A similar calculation can be performed on a FQH system at filling factor $\nu=1/3$ on a torus. In that case, we use the threefold groundstate of the pseudopotential hamiltonian\cite{Haldane-1983PhysRevLett.51.605} that generates the Laughlin state. We can then change the aspect ratio $\alpha=L_y / L_x$ of the torus to reach the thin torus limit and see the effect on the PES. Such results are shown in Figs. \ref{entspectrum}b, \ref{entspectrum}d and \ref{entspectrum}f. We clearly that both the FCI and the FQH PES exhibit the same structure. We stress that the PES of the FQH in the thin torus limit (almost reached in Fig. \ref{entspectrum}b) corresponds to the one of a CDW.

\section{Thin torus limit in the FQH effect}

The physics of the exact flat-band $1$-D models presented above is identical to that of the thin-torus limit of the FQH. Thus it is highly relevant to look back at this case and see which emerging similar structure could be look after in FCI.

\subsection{Hierarchy of bands}

\begin{figure}[htb]
\includegraphics[width=0.49\linewidth]{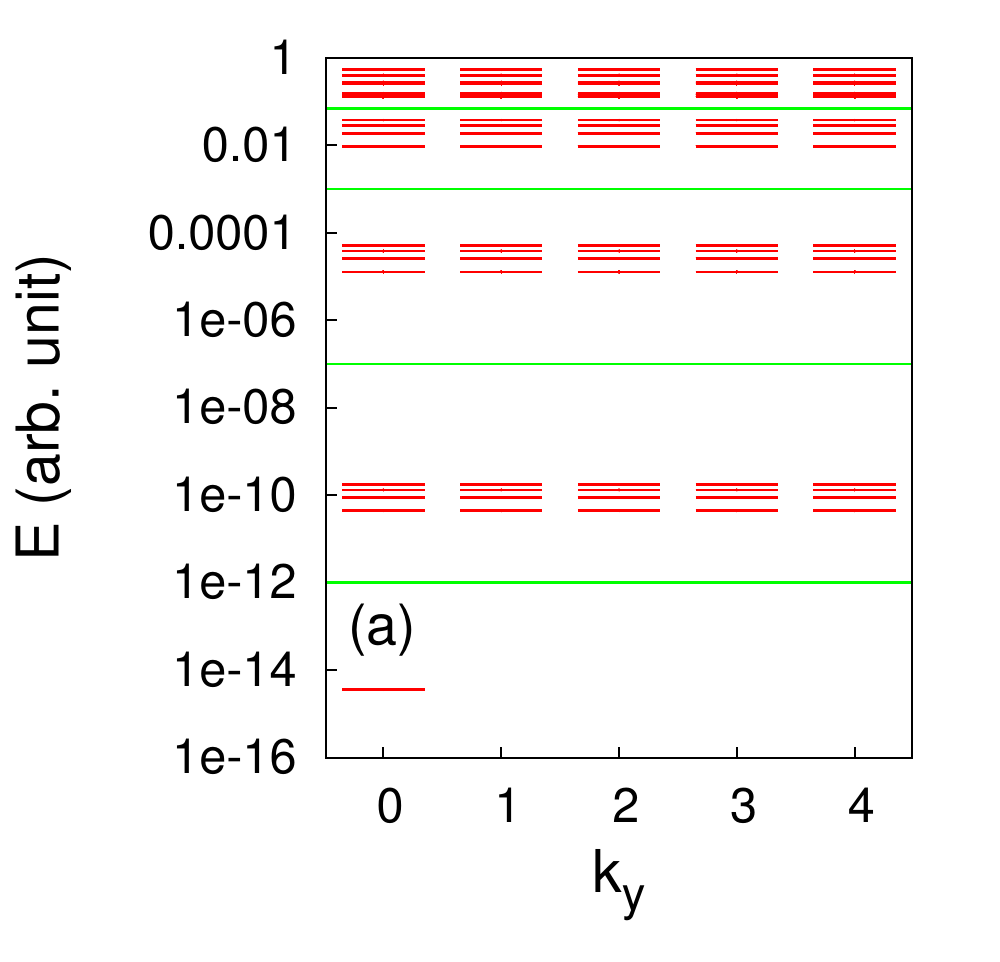}
\includegraphics[width=0.49\linewidth]{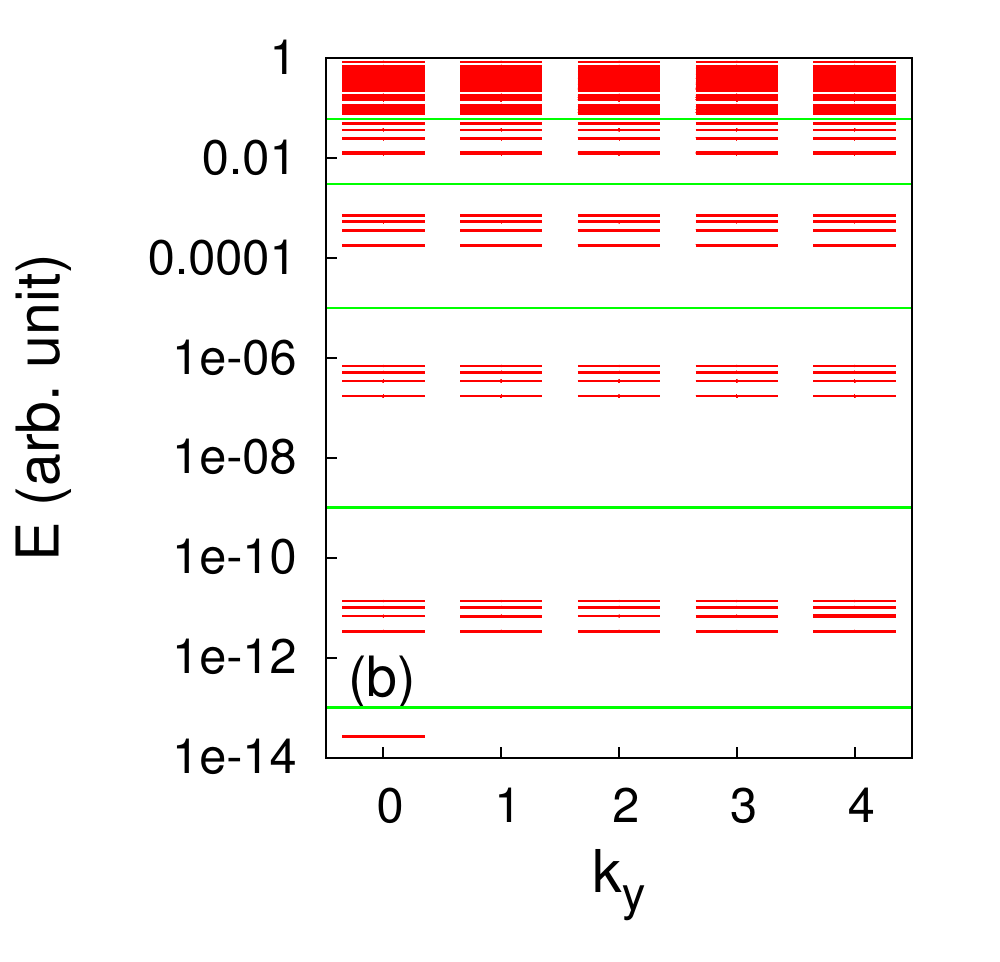}
\caption{{\it Left panel :} Energy spectrum for FQH on torus and the Laughlin $\nu=1/5$ model interaction with $N_e=5$ electrons and aspect ratio $\alpha=0.07$. {\it Right panel :} Similar energy spectrum for the Laughlin $\nu=1/7$ model interaction with $N_e=5$ electrons and aspect ratio $\alpha=0.095$. In both cases we observe a zero manifold and several bands associated to the $(1,r)$ exclusion principle. The absence of a gap related to the $(1,2)$ for the $\nu=1/7$ case is an artefact of the limited numerical accuracy that does not allow to consider smaller aspect ratio without collapsing the $(1,7)$ gap. The plots do not  show  the $5$ and $7$ fold center of mass momentum degeneracy, as the FQH spectrum is exactly degenerate,}\label{laughlinm}
\end{figure}

The thin-torus limit of Haldane pseudo-potential Hamiltonians is that of electrostatic interactions (or the generalized $k$-body interaction for the RR state). These are just density-density (in the case of $Z_k$ RR state, we have a $k$-body density$^k$) interactions of density operators of the torus \emph{momentum} $j =0 \ldots N_\Phi-1$. For example, for a torus of length $L_x, L_y$ (for simplicity we chose rectangular torus), $L_x L_y= 2 \pi N_\phi$ with LLL orbitals in the Landau gauge $A_y = B x$ are indexed by a momentum quantum number $j$:
\beq
\phi_j(z) = \frac{1}{\sqrt{\pi^{1/2} L_y}}e^{- x^2/2} \sum_{k \in {\cal{Z}}} e^{- L_x (k + \frac{j}{N_\phi})z - L_x^2 (k + \frac{j}{N_\phi})^2/2}  \label{thintoruslimitFQH}
\eneq The two-body Haldane $1/3$ pseudopotential Hamiltonian $V(r)= \nabla^2 \delta({\bf{r}})$ projected to the LLL,   which has the Laughlin $\nu=1/3$ state as the zero energy ground-state, can be written using the LLL orbitals in second quantized form as $\sum_{j_{1,2,3,4}} V_{j_1, j_2, j_3, j_4} c_{j_1}^\dagger c_{j_2}^\dagger c_{j_3} c_{j_4}$. In  the isotropic limit $V_{j_1, j_2, j_3, j_4} $ can be expressed as an infinite sum \cite{chakraborty1995quantum} of exponentials, but has large matrix elements for large separations $j_1- j_2$, etc.  It also does not assume any particularly nice form and cannot be written as a density-density interaction, i.e. there is no sense in which $V_{j_1, j_2, j_3, j_4}$ with $j_2=j_3, j_1=j_4$ is larger than other matrix elements. In the thin torus limit $L_y\rightarrow 0$ (better described by the ratio $\alpha=L_y/L_x \rightarrow 0$, with $L_x L_y = 2 \pi N_\phi$ constant), a re-organization of the matrix elements takes place and several things occur. First, the interaction becomes only density-density,  and the matrix elements with $j_2 \ne j_3, j_1\ne j_4$ are exponentially suppressed. Amongst the density-density interactions, the short-range ones have the highest weight and longer range density density interactions decay exponentially.  A short calculation gives the matrix elements for the Haldane $1/3$ pseudopotential Hamiltonian in terms of the aspect ratio $\alpha$ and number of fluxes $N_\phi$ to be:
\beq
V = \sum_{j=1}^{N_\phi-1} \frac{2\pi}{ \alpha N_\phi} e^{- \frac{\pi}{ \alpha N_\phi}} n_j n_{j+1} + \frac{8 \pi}{ \alpha N_\phi} e^{- \frac{4 \pi}{ \alpha N_\phi} } n_{j} n_{j+2}     \label{thintoruslimitFQHhamiltonian}
\eneq where $n_j = c_j^\dagger c_j$ is the density of the $j$'th orbital momentum. This is an exactly solvable model. The exact zero modes of this Hamiltonian are the Slater determinants with occupation number configurations  given by the rule that no electron can be closer than 3 orbitals to another electron (if it is, one or both of the above potentials are nonzero). This rule is identical to a generalized Pauli principle which postulates that there should be no more than $1$ particle in $3$ consecutive orbitals (we call this $(1,3)$ Pauli principle). At filling $N_e/N_\phi= 1/3$, there are $3$ wavefunctions (which are non-interacting Slater determinants) given by the configurations $100100100100\ldots$, $010010010010\ldots$ and $001001001001$. The only difference between the zero modes of our flat band model and those of the FQH thin torus limit is that they are CDW in real space in the flat band model and in momentum space in the FQH. What are the excitations of the FQH at fillings below $1/3$? Notice the separation of scales for $r\rightarrow 0$ in the thin torus Hamiltonian \ref{thintoruslimitFQH}: in this limit, the $ n_{j} n_{j+2} $ interaction has a coefficient  $4 e^{- \frac{3 \pi}{ \alpha N_\phi} }$ exponentially smaller than that of the $n_j n_{j+1}$ interaction. We hence expect that the full Hamiltonian has 3 bands of states. First the zero modes. Then the first band of excitations above the zero-modes of  \ref{thintoruslimitFQHhamiltonian} who will be zero modes of the $n_j n_{j+1}$ interaction (which has a large penalty). The wavefunctions of these excitations are Slater determinants  made out of configurations which satisfy the Pauli principle that there should be no more than $1$ particle in $2$ consecutive orbitals (we call this $(1,2)$ principle). The energy of the excitation spectrum of the band of states described by this principle starts at $\frac{8 \pi}{ \alpha N_\phi} e^{- \frac{4 \pi}{ \alpha N_\phi} }$ and ends at $(N_e-1) \frac{8 \pi}{ \alpha N_\phi} e^{- \frac{4 \pi}{ \alpha N_\phi} }$ where $N_e$ is the number of electrons (we work at filling $\le 1/3$, so the zero modes are the FQH  quasiholes). For the FQH, the upper energy level of the $(1,2)$ excitations is (for any number of electrons) smaller than the lower level of the  highest energy band: $(N_e-1) \frac{8 \pi}{ \alpha N_\phi} e^{- \frac{4 \pi}{ \alpha N_\phi} } <<  \frac{2\pi}{ \alpha N_\phi} e^{- \frac{\pi}{ \alpha N_\phi}}$. Hence the $(1,2)$ band of excitations (see Fig. \ref{gsspectrum}b) is separated by a gap $ \frac{8 \pi}{ \alpha N_\phi} e^{- \frac{4 \pi}{ \alpha N_\phi} } $ from the zero modes and by a gap  $\frac{2\pi}{ \alpha N_\phi} e^{- \frac{\pi}{ \alpha N_\phi}}- (N_e-1) \frac{8 \pi}{ \alpha N_\phi} e^{- \frac{4 \pi}{ \alpha N_\phi} } $ from an upper band of excitations. The upper band of excitations now is made out of all remaining fermionic (Pauli principle $(1,1)$) excitations less  the ones in the $(1,2)$ and $(1,3)$ bands. The Hamiltonian Eq[\ref{thintoruslimitFQHhamiltonian}] has other terms,  including ones that cannot be written out as density-density interactions, but they are all at least a factor $e^{- 2 \frac{\pi}{ \alpha N_\phi}} $ smaller than the largest term in $V$.

\begin{figure}[htb]
\includegraphics[width=0.48\linewidth]{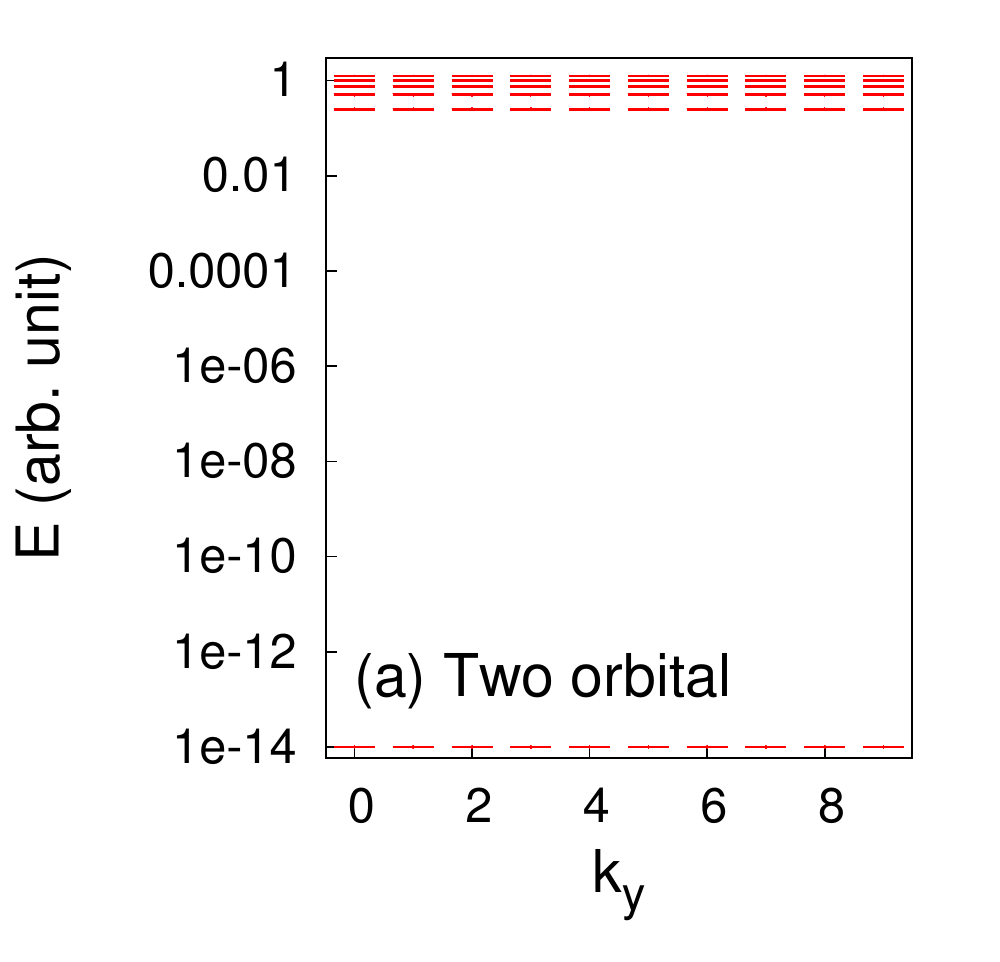}
\includegraphics[width=0.48\linewidth]{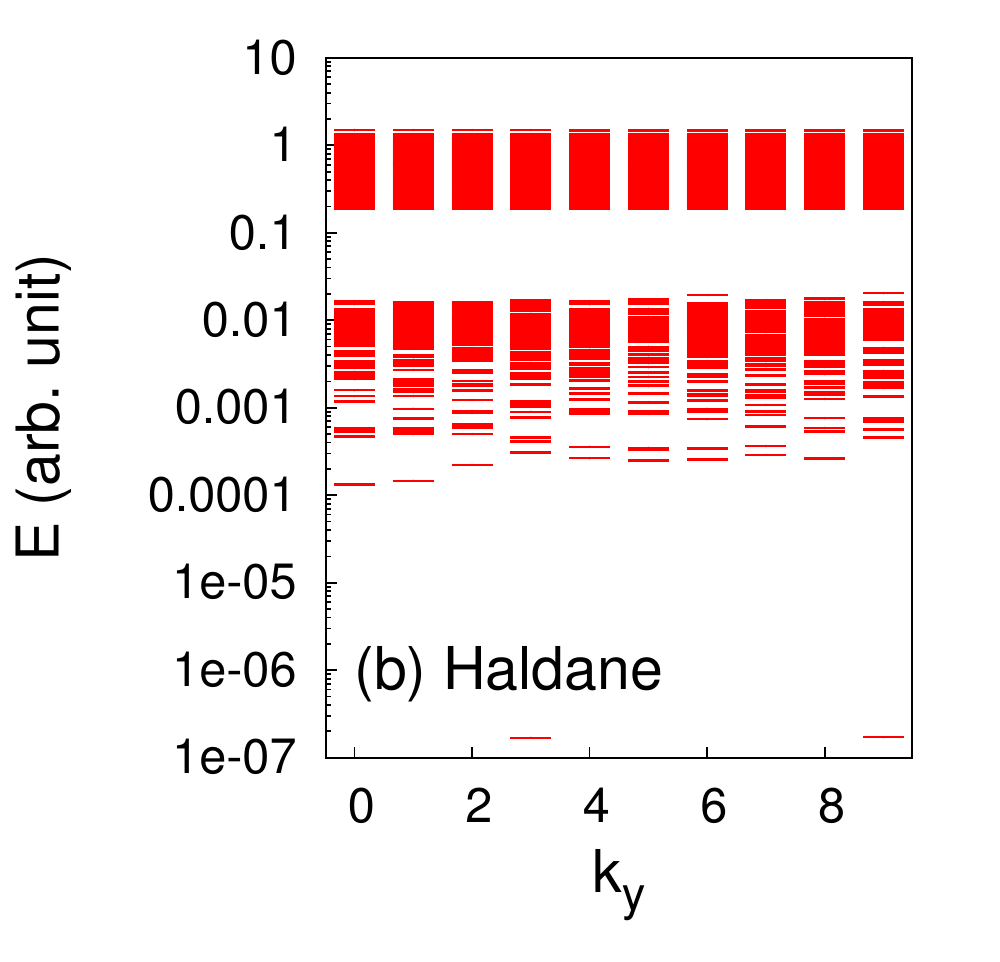}

\includegraphics[width=0.48\linewidth]{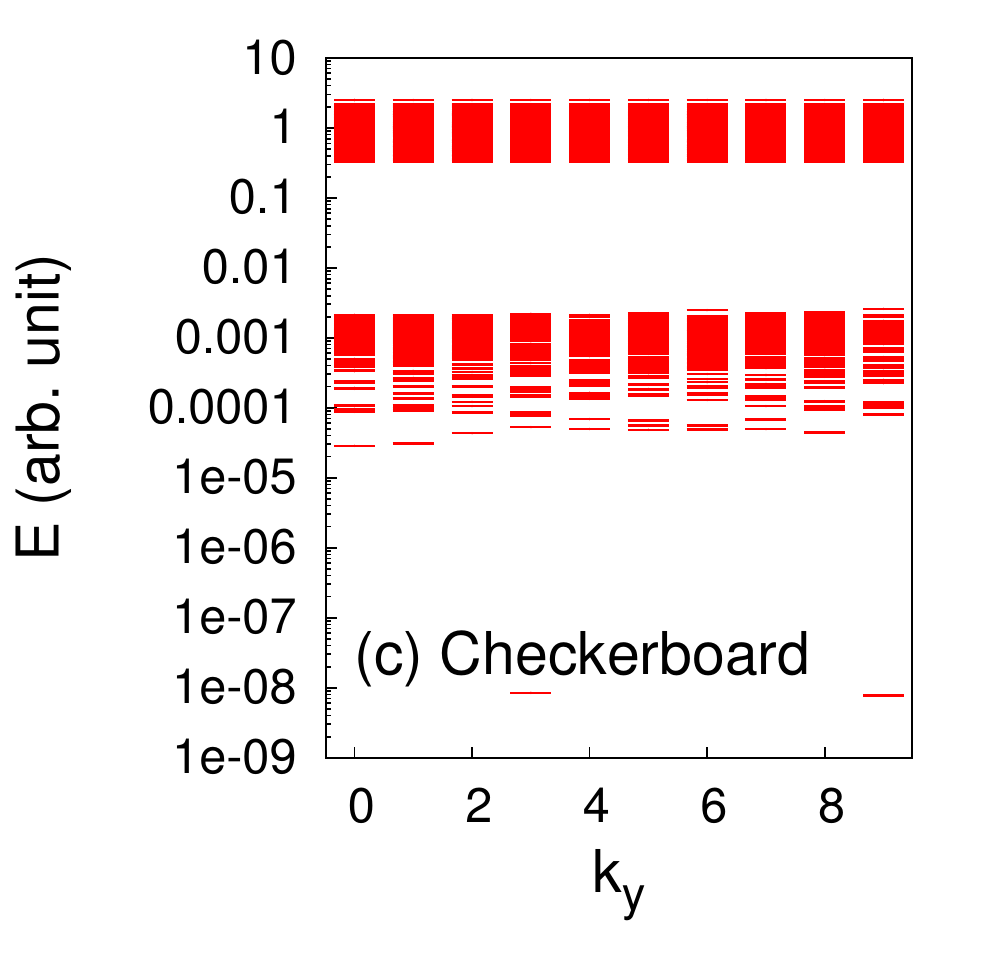}
\includegraphics[width=0.48\linewidth]{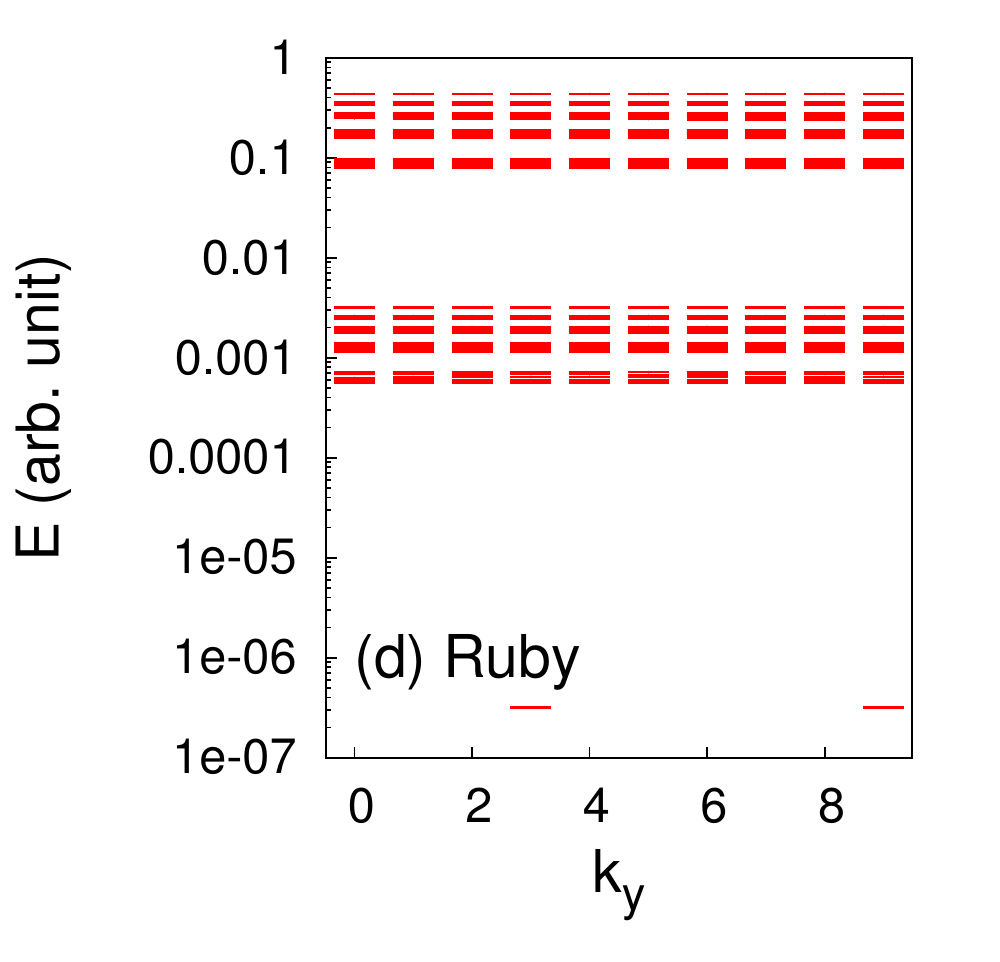}
\caption{Energy spectra for $N=6$ fermions with $18$ unit cells for different models in the thin torus limit $(N_x=1)$. The four models are : the two orbital model with $V_1=0$ (a), the Haldane model (b), the checkerboard lattice model (c) and the ruby lattice model (d). Only the momentum sectors with $k_y \le 9$ are shown, the other momentum sectors are related to these ones by the inversion symmetry. Notice that the two orbital models has a highly degenerate manifold for $V_1=0$.}\label{fcispectra}
\end{figure}

This hierarchy of bands in the thin torus limit is generic to the pseudopotentials in FQH, and despite a literature search in this extensively researched subject, we were not able to find previous mention of it. This statement seems to be valid any Laughlin states $\nu=1/m$ where we observe $(1,m)$ zero modes and then $(1,m-1)$ up to $(1,1)$ bands, as depicted in Fig. \ref{gsspectrum}b for $m=3$, in Fig. \ref{laughlinm}a for $m=5$, and in Fig. \ref{laughlinm}b for $m=7$. 

\subsection{Numerical analysis of FCI models}

\begin{figure}[htb]
\includegraphics[width=0.98\linewidth]{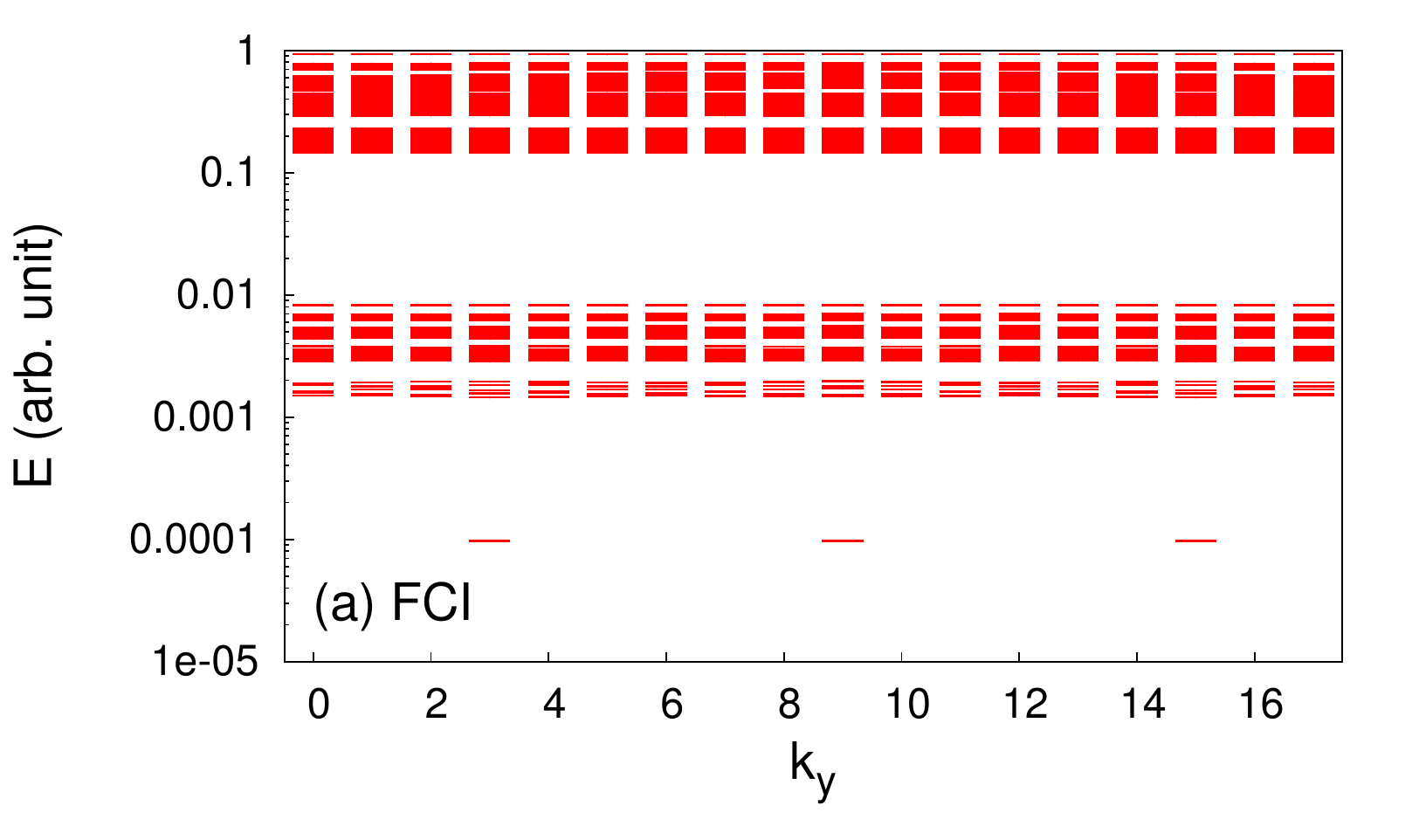}
\includegraphics[width=0.98\linewidth]{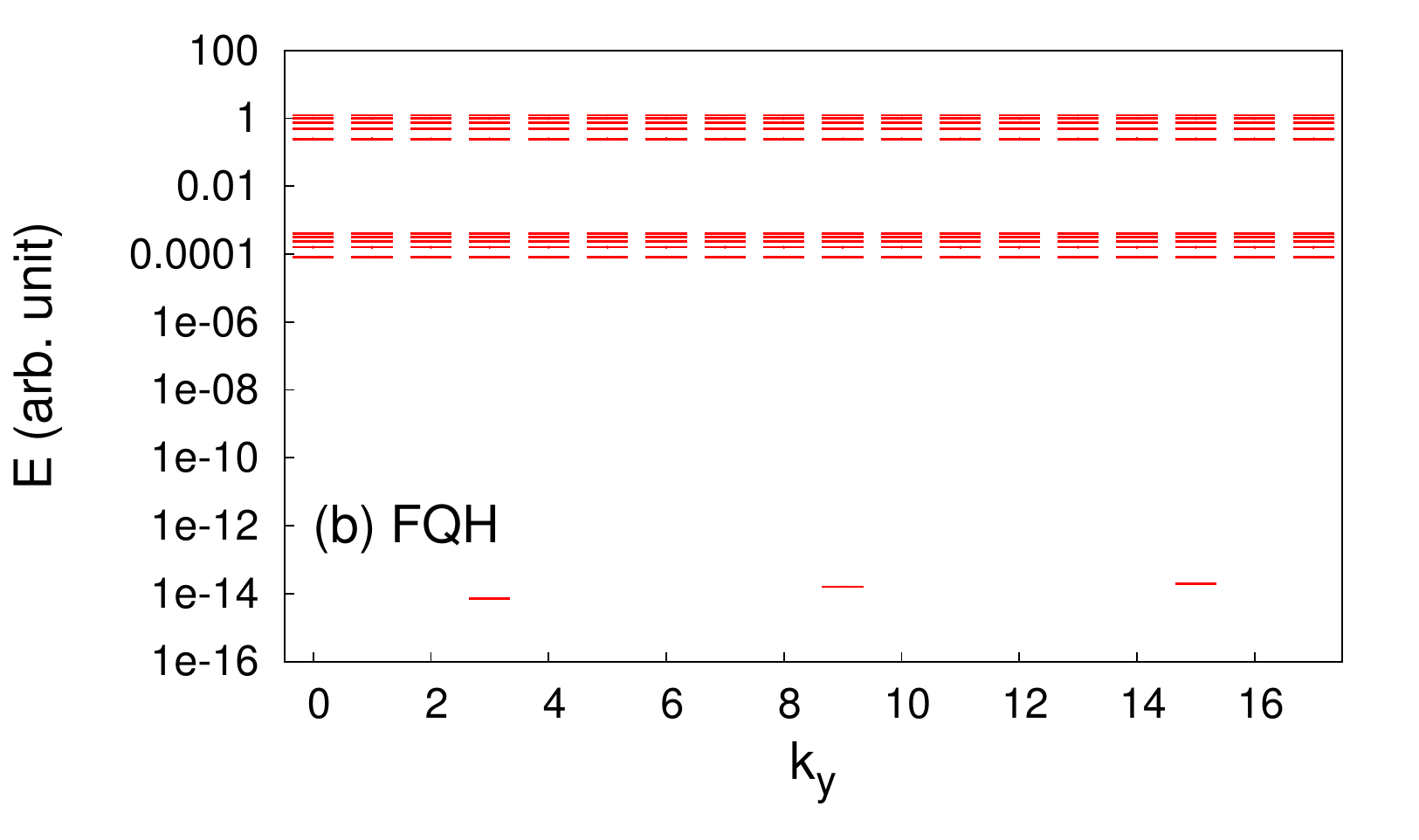}
\caption{Energy spectra for $N=6$ fermions for the Kagome lattice with $18$ unit cells (upper panel) and FQH on the torus geometry with $N_{\Phi}=18$ flux quanta and  aspect ratio $1/18$ (lower panel). In the FQH case, we only display the momentum sectors that cannot be related by the center of mass degeneracy. The small dispersion of the low energy manifold on the FQH spectrum is due to the numerical accuracy.}\label{gsspectrum}
\end{figure}

\begin{figure}[htb]
\includegraphics[width=0.98\linewidth]{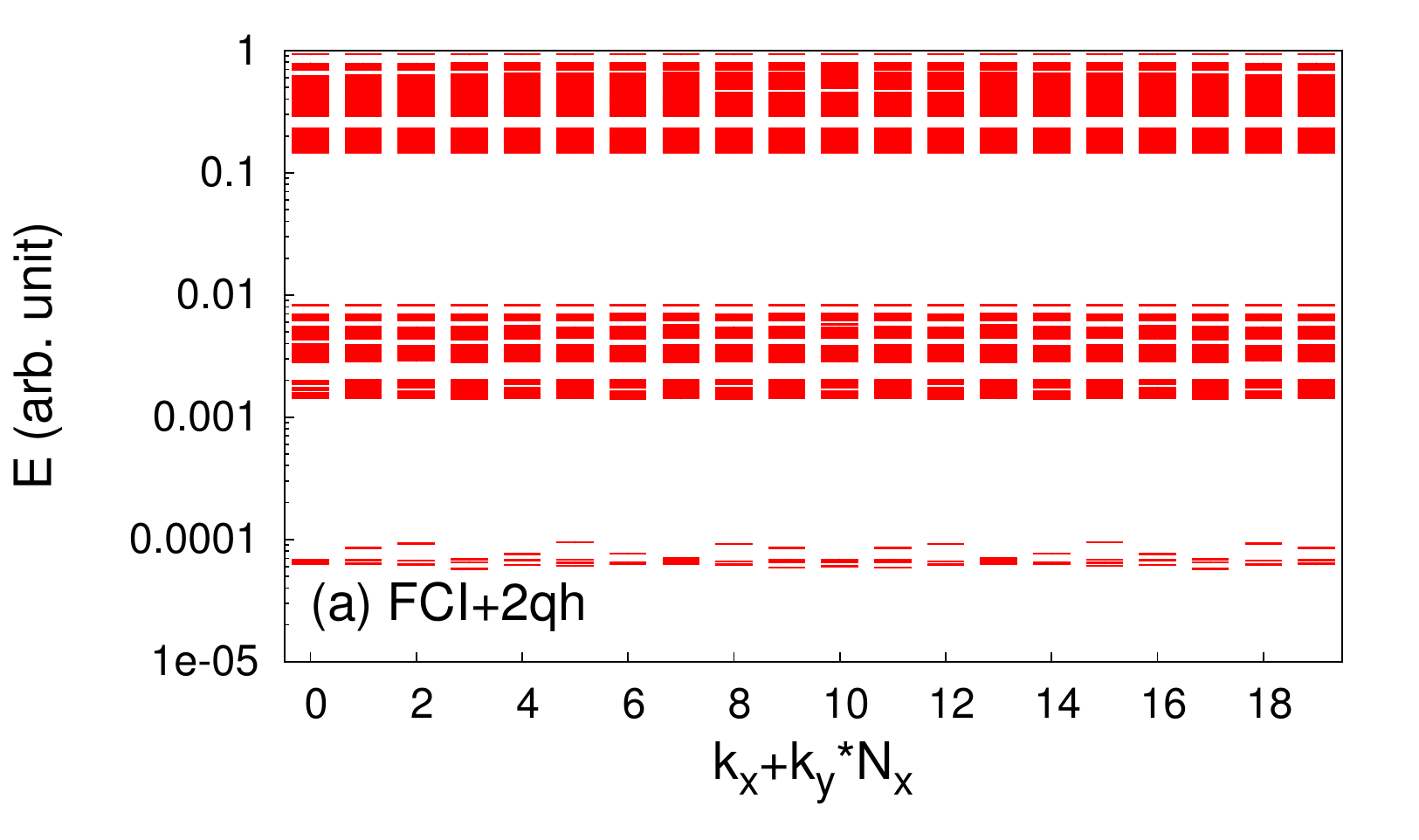}
\includegraphics[width=0.98\linewidth]{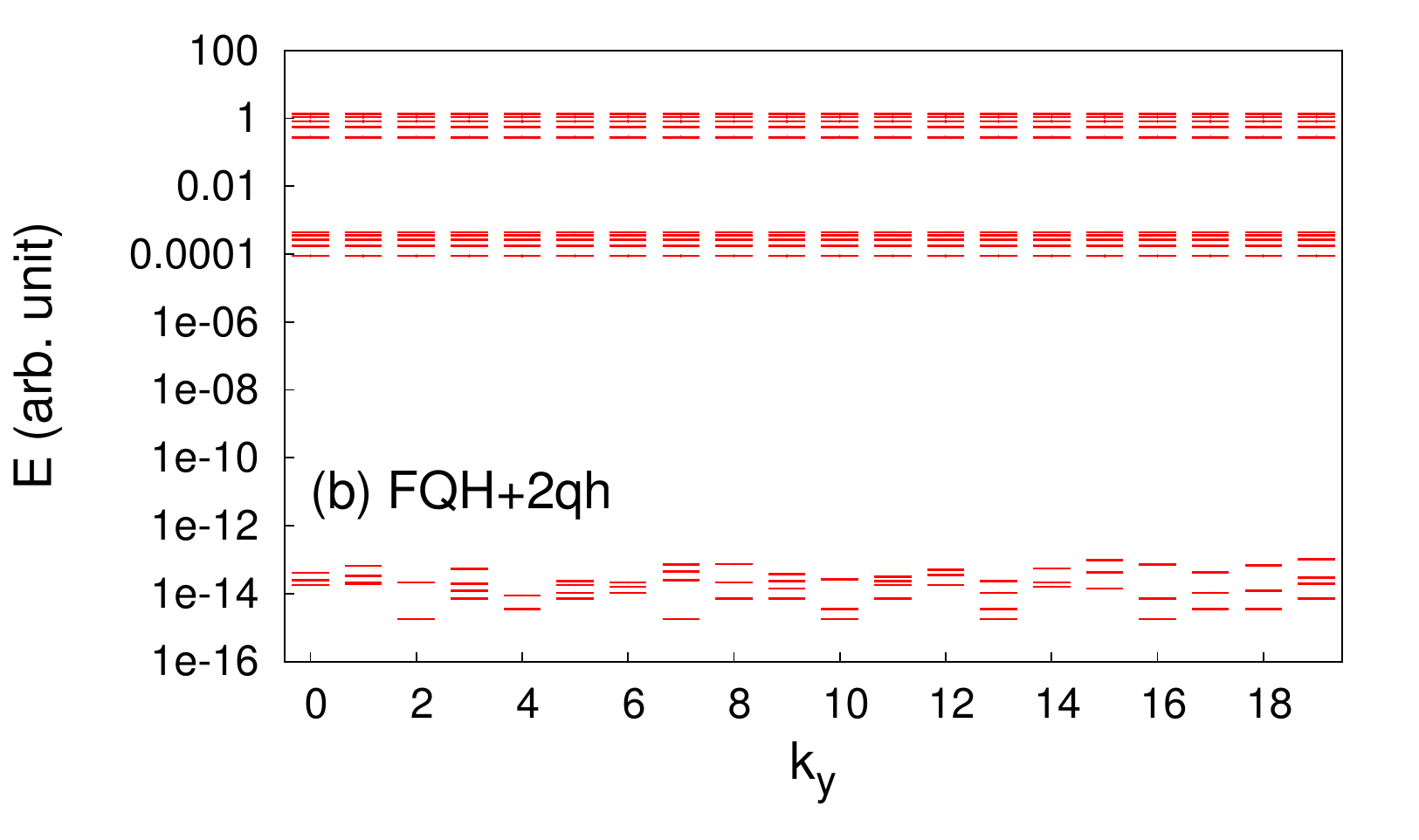}
\caption{Energy spectra for $N=6$ fermions for the Kagome lattice with $20$ unit cells (upper panel) and FQH on the torus geometry with $N_{\Phi}=20$ flux quanta and aspect ratio $1/20$ (lower panel). This situation corresponds to the Laughlin state with two added quasihole. In the FQH case, we only display the momentum sectors that cannot be related by the center of mass degeneracy. The small dispersion of the low energy manifold on the FQH spectrum is due to the numerical accuracy.}\label{qhspectrum}
\end{figure}

We now perform a numerical analysis of the thin torus limit of the Kagome model, which, in its isotropic limit, has the strongest FCI state. We find a clear separation of the states described by the Pauli principles $(1,3)$, $(1,2)$, and $(1,1)$ in both the case $N_e/N_y=1/3$ (see Fig[\ref{gsspectrum}]a). A similar result is also observe for $N_e/N_y <1/3$, i.e. when quasiholes are considered (see Fig[\ref{qhspectrum}]a).

We now conjecture that good FCI states have, in the thin torus limit, a similar separation of scales as the FQH. The FCI projected Hamiltonian is similar in spirit to the FQH hamiltonian: when projected, in the isotropic limit, it cannot be written as density-density interactions of any range. In the thin torus limit, at least for models where the band is flat without the need for long-range projectors, the FCI projected Hamiltonian becomes short-ranged and writable in terms of just density-density interactions For FCI Hamiltonians which when projected in the thin torus limit, exhibit a separation of scales similar to that in the FQH we observe strong FCI states in the isotropic limit. This is the criterion which we conjecture is essential in finding FCI states on the lattice, and we support it below with numerical data. Fig. \ref{fcispectra} shows the energy spectrum for several models. The Ruby\cite{hu-PhysRevB.84.155116,Wu-2012PhysRevB.85.075116}, checkerboard\cite{sun-PhysRevLett.106.236803,sheng-natcommun.2.389,neupert-PhysRevLett.106.236804,regnault-PhysRevX.1.021014}, and Haldane model\cite{haldane-PhysRevLett.55.2095,Wu-2012PhysRevB.85.075116},  which exhibit decreasingly strong FQH states in the isotropic limit for fermions, have a decreasingly clear separation of scales in the thin torus limit. The ruby model, with one of the strongest FQH Laughlin state \cite{Wu-2012PhysRevB.85.075116} has a clear manifold of $3$ zero modes (only $2$ are shown, the third one being the inversion symmetric of the one at $k_y=3$, and shows up at $k_y=15$), then a manifold of $(1,2)$ Pauli principle states clearly separated from both the zero modes and from the highest  energy manifold of $(1, 1)$ states. The bandwidth of each of these manifolds is small, and they are clearly defined. The Haldane model still shows a similar separation of scales, but the bandwidth of each manifold is now large and unlike the FQH. The two-orbital model with just $V_0$ does not show such a separation (for example, there is no difference between the $(1,3)$ and $(1,2)$ modes which appear at zero energy), as we can prove from analytics (we need the $V_1$ to lift the degeneracy of $(1,2)$ vs $(1,3)$ modes), and the isotropic state in the two orbital model is not a Laughlin state at this lattice size.

\section{Conclusion}

In this paper we have analyzed the thin torus limit of several models which are supposed to exhibit a FCI state. The thin torus limit of lattice models was defined as the $N_x=1$ limit of an $N_x\times N_y$ site lattice of electrons in the presence of any Hubbard-like interaction. We found a series of one-body models with finite-range hoppings which for isotropic aspect ratio exhibit nonzero Chern numbers in the nontrivial side and which, in the thin-torus limit, exhibit perfectly flat bands. In the thin torus limit, the nontrivial side of the  insulator exhibits dimerization, trimerization, or more generally $n$-merization etc. Our thin torus limit one-body Hamiltonians can also be thought as inversion symmetric one-dimensional insulators with half charge polarization on the edge. In our toy models, due to the short-range form of the projection operator, we can diagonalize exactly Hubbard-type interactions in the lowest band. We show that projecting the interaction to the lowest band involves an increase in its range by exactly the distance of the $n$-merized bond in our one-body model. In our toy models the interactions can be tuned so that when projected they give exactly the correct ground-state degeneracy for the state we are considering. We choose the interactions so that the spectrum matches qualitatively that of the thin-torus limit of the FQH. Due to their simplicity,  we can exactly solve the full spectrum of such interactions, and show that it separates into bands of states spanned by Slater determinants in real space satisfying a certain exclusion statistics hierarchy. For example, for the interaction responsible for the $1/3$ filling Laughlin state, in the thin torus limit the spectrum is made out of $3$ bands: the zero modes are exactly the ground-state and quasiholes of the Laughlin state satisfying a Pauli $(1,3)$ principle that disallows more than $1$ particle in $3$ consecutive orbitals. The middle band satisfies a $(1,2)$ Pauli principle, while the highest energy band satisfies a $(1,1)$ (usual fermionic) Pauli principle. In the thin torus limit, all these states are CDW (or its excitations), which shows pure counting of ground-states and excitations is not enough to fully determine the topological nature of a state. We then compute analytically  the entanglement spectrum of the degenerate manifold of states and show that the entanglement spectrum of the CDW state and that of the isotropic FCI state differ considerably, both qualitatively and quantitatively, and can be used to differentiate between the states. We identify the CDW and FCI structures in the entanglement spectrum and show how the CDW spectrum evolves into that of the FCI upon going from the thin-torus to the isotropic limit.  We then numerically analyze several FCI models and present numerical evidence for the conjecture that strong isotropic FCI states appear when there exists a separation of scales in the thin torus limit similar to that of the FQH.

\emph{Acknowledgements}
We thank Z. Papic, Yangle Wu, A. Alexandradinata, B. Estienne and F.D.M. Haldane for useful discussions.  BAB was supported by Princeton Startup Funds, NSF CAREER DMR-095242, ONR - N00014-11-1-0635, Darpa - N66001-11-1-4110, Packard Foundation and Keck grant. NR was supported by  NSF CAREER DMR-095242, ONR - N00014-11-1-0635, Packard Foundation and Keck grant.

\bibliography{ThinTorusLimit.bib}

\end{document}